\documentclass[10pt,letterpaper]{article}
\usepackage[top=0.85in,left=2.75in,footskip=0.75in]{geometry}
\usepackage{graphicx}
\usepackage{authblk}
\usepackage{amsmath}
\usepackage{amsfonts}
\usepackage{amssymb}
\usepackage{algorithm,algorithmic}
\usepackage{multirow}
\usepackage{setspace}
\usepackage{appendix}
 \usepackage{vcell}
\usepackage{url}

\usepackage{changepage}

\usepackage[utf8x]{inputenc}

\usepackage{textcomp,marvosym}

\usepackage{cite}

\usepackage{nameref,hyperref}

\usepackage[right]{lineno}

\usepackage{microtype}
\DisableLigatures[f]{encoding = *, family = * }

\usepackage[table]{xcolor}

\usepackage{array}

\newcolumntype{+}{!{\vrule width 2pt}}

\newlength\savedwidth

\raggedright
\usepackage[aboveskip=1pt,labelfont=bf,labelsep=period,justification=raggedright,singlelinecheck=off]{caption}

\makeatletter
\renewcommand{\@biblabel}[1]{\quad#1.}
\makeatother

\usepackage{lastpage,fancyhdr,graphicx}
\usepackage{epstopdf}
\fancyheadoffset[L]{2.25in}
\fancyfootoffset[L]{2.25in}
\newcommand{\Name}{{\tt MUPen2DTool}}

\begin{document}
\vspace*{0.2in}

\begin{flushleft}
{\Large
\textbf\newline{\Name: a Matlab Tool for 2D Nuclear Magnetic Resonance relaxation data inversion} 
}
\newline
\\
Villiam Bortolotti\textsuperscript{1\Yinyang\ddag},
Leonardo Brizi\textsuperscript{2\ddag},
Germana Landi\textsuperscript{3\Yinyang},
Anastasiia Nagmutdinova\textsuperscript{1\ddag},
Fabiana Zama\textsuperscript{3\Yinyang*}
\\
\bigskip
\textbf{1} Department of Civil, Chemical, Environmental, and Materials Engineering, University of Bologna, Italy
\\
\textbf{2} Department of Physics and Astronomy, University of Bologna, Italy
\\
\textbf{3} Department of Mathematics, University of Bologna, Italy
\\
\bigskip
%
%
\Yinyang These authors contributed equally to mathematical modelling and development of numerical methods.

\ddag These authors contributed equally to the development of the software interface and data acquisition.




* corresponding author: fabiana.zama@unibo.it

\end{flushleft}
\section*{Abstract}
Accurate and efficient analysis of materials properties 
from Nuclear Magnetic Resonance (NMR) relaxation data requires
robust and efficient inversion procedures.
Despite the great variety of applications requiring to process two-dimensional NMR data (2DNMR), a few software tools are freely available.
The aim of this paper is to present \Name\, an open-source MATLAB based software tool for 2DNMR data inversion.
The user can choose among several types of NMR experiments, and the software provides codes that can be used and extended easily. Furthermore, a  MATLAB interface makes it easier to include users own data.
The practical use is demonstrated in the reported examples of both synthetic and real NMR data. 
\nolinenumbers

\section*{Introduction}
Nuclear Magnetic Resonance relaxometry of $^{1}$H nuclei ( $^{1}$H NMR) can give crucial information about the properties of many materials, ranging from cement \cite{anas2021} to biological tissues \cite{fanta2011}.
So, for example in case of porous media, $^{1}$H NMR permits the accurate estimate of important petrophysical parameters, such as porosity, saturation and permeability. For example, borehole $^{1}$H NMR is extensively used in oil and gas reservoir characterisation, and recent technological advances have led to tools suitable for environmental applications (see details in \cite{USGS}).
Usually the NMR parameters investigated are relaxation times (longitudinal T$_{1}$ and/or transverse T$_{2}$) and self-diffusion coefficient (D) as they are sensitive to the local physical environment and can also provide some chemical information.
As there may be a range of the NMR parameter values that characterize a given system, in order to correctly interpret NMR experiment results and compute the distributions of these parameters, it is necessary to processes the experimental data with a robust and accurate inversion procedure.
Recently, two-dimensional NMR (2DNMR) techniques are gaining increasing importance in analysing different porous media \cite{mitc2012}. Moreover, new computationally intensive applications, such as multidimensional logging and general 3DNMR data inversion \cite{bsun2005}, require efficient methods for the inversion of 2DNMR data. For all these reasons, there is an increasing request of software that can be easily applied to process 2DNMR data to compute 2D parameter distribution (NMR maps).
%
Even if a considerable amount of inversion methods has been proposed in literature, the software tools implementing such methods are seldom freely available for testing. 
%
%
We strongly believe that open source software gives a determinant contribution to the progress of knowledge by making it possible for scholars to compare and improve their achievements. Therefore, starting from 2009,  we released {\tt Upenwin} \cite{UPwin} and recently {\tt Upen2dTool} \cite{SoftX}, an open-source software tool for 2DNMR data inversion. 
\\
From a mathematical point of view, the problem of computing the two-dimensional relaxation time distributions from NMR data is a linear ill-posed problem modelled by a Fredholm integral equation with separable kernel. The strong ill-conditioning and the presence of data noise make the inverse problem very challenging. A regularisation technique is applied to reformulate the inversion problem  as a non-negatively constrained optimisation problem, whose objective function  contains a data fitting term and a regularisation term.
In 	\cite{Borgia1998, Borgia2000} the Uniform Penalty (UPEN) principle, a multiple-parameters locally adapted Tikhonov-like regularization method, has been stated for one-dimensional NMR data and has been implemented in {\tt Upenwin} software. Successively, in 2016, such principle has been extended for two-dimensions data \cite{Bortolotti2016}, and further analysed and improved in \cite{ncmip17, micromeso17}. In 2019, {\tt Upen2dTool} has been made available \cite{SoftX}.
\\
Both {\tt Upenwin} and {\tt Upen2dTool} implement a $L2-$norm locally adapted regularisation (in one and two dimensions, respectively), where the automatic computation of the regularisation parameters follows the UPEN principle. Although such methods compute very accurate distributions, their computational cost may be high since they require the solution of several non-negatively constrained least-squares problems. For this reason, a new method has been studied and proposed in \cite{ComputGeoSci2021} which represents a substantial change in the inversion strategy, consisting of adding an $L1$ penalty term to the locally adapted $L2$ term and removing the non-negativity constraint. The new method allows us to substantially improve the computation efficiency of the inversion process and for some kind of 2D data, to obtain even more robust and accurate NMR maps. 
We now release the Multiple Uniform Penalty 2D Tool (\Name) open source software implementing the method proposed in \cite{ComputGeoSci2021}. \Name\  consists of the source code, software documentation and a user guide  which contains an installation guide, a technical description of synthetic NMR tests and input data format.
\Name \ comes also with a user friendly GUI (Graphical User Interface) that guides the user in the different steps of the inversion process. Moreover  NMR data of several representative examples are available to help the interested user to assess the toolbox efficiency and effectiveness.
The GUI makes it easy to handle the inversion parameters and inspect the loaded data, therefore our software can be flexibly used in the analysis of different types of samples. 
Being \Name\ open source and user friendly, we believe that it has a large number of potential users from different application fields.

In this paper we describe the software through a brief overview of the implemented algorithms and a detailed analysis of the 2D distributions computed from the data set enclosed in the software package.

The paper has the following structure.
We first introduce the general structure of  \Name \ describing its key features. Then, we present the problem of NMR data inversion and the characteristics of the implemented regularization algorithm.  Finally, we report the software validation on several representative NMR relaxometry data relative to different types of samples.
\section*{The problem of NMR data inversion\label{method}}
%
%
In this section, we describe the mathematical model for NMR data inversion and the numerical scheme used by \Name \ for its solution.
\subsection*{The continuous model}
In \Name, we consider 2DNMR maps corresponding to $T_1$-$T_2$, $T_2$-$T_2$ and $T_2$-$D$ relaxation data; in all these cases, the measured NMR signal is supposed to be related to an underlying distribution function by a Fredholm integral equation of the first kind. 

\emph{$T_1$-$T_2$ case.}
In a conventional Inversion-Recovery (IR) or Saturation Recovery (SR) experiment detected by a Carr-Purcell-Meiboom-Gill (CPMG) pulse
sequence \cite{Blumich2005}, the relaxation data $S(t_1, t_2)$ depending on $t_1$, $t_2$ evolution times can be expressed as:
\begin{equation}\label{eq:modelT1T2}
  S(t_1, t_2) = \iint_{0}^{\infty} k_1(t_1, T_1)k_2(t_2, T_2)F(T_1, T_2) dT_1 dT_2 + e(t_1, t_2)
\end{equation}
where  $F(T_1, T_2)$ is the unknown distribution of $T_1$ and $T_2$ relaxation times and the kernels $k_1$ and $k_2$ have the expression
\begin{equation}
   k_1(t_1,T_1)=\left\{
  \begin{array}{ll}
    1-2\exp(-t_1/T_{1}), & \hbox{for IR sequence} \\
    1-\exp(-t_1/T_{1}),  & \hbox{for SR sequence}
  \end{array}
\right. \text{ and }
\ \ k_2(t_2,T_{2})=\exp(-t_2/T_{2}).
\end{equation} 

Here and henceforth, the function $e(\cdot , \cdot)$ represents Gaussian additive noise.

\emph{$T_2$-$T_2$ case.}
In a CPMG-CPMG experiment, the measured data $S(t_1, t_2)$ is related to the underlying distribution $F(T_{21}, T_{22})$ by the integral equation
\begin{equation}\label{eq:modelT2T2}
  S(t_1, t_2) = \iint_{0}^{\infty} k_1(t_1, T_{21})k_2(t_2, T_{22})F(T_{21}, T_{22}) dT_{21} dT_{22} + e(t_1, t_2)
  \end{equation}
where both kernels $k_1$ and $k_2$ refer to transversal relaxation times $T_{21}$, $T_{22}$ and are defined as 
\begin{equation}\label{}
  k_1(t_1,T_{21})= \exp(-t_1/T_{21}),  \ \ \ k_2(t_2,T_{22})=\exp(-t_2/T_{22}).
\end{equation} 

\emph{Diffusion-$T_2$ case.}
In a Stimulated Echo-CPMG experiment, the acquired echo amplitude $S(t_1,t_2)$ can be expressed as
\begin{equation}\label{eq:modelT2D}
  S(t_1, t_2)=\iint_{0}^{\infty} k_1(t_1, T_2)k_2(t_2,D)F(T_2,D) dT_2 dD + e(t_1,t_2)
\end{equation}
where the kernels $k_1$ and $k_2$ are
\begin{equation}\label{}
  k_1(t_1,T_2)= \exp(-t_1/T_2),  \ \ \ k_1(t_2,D)=\exp(- t_2 \cdot D)).
\end{equation} 
%

\medskip
The objective is to estimate the $T_1$-$T_2$, $T_2$-$T_2$ or $D-T_2$ map from the measured data; this
inversion is an ill-posed problem, which means that small noise in the data can cause significant changes in the computed 2D distribution.
\subsection*{The discrete model}
The discretization of the linear integral equations \eqref{eq:modelT1T2}, \eqref{eq:modelT2T2} and \eqref{eq:modelT2D} leads to the liner system 
\begin{equation}\label{mod1}
    \mathbf{K} \mathbf{f} + \mathbf{e}= \mathbf{s}
\end{equation}
where $\mathbf{K}= \mathbf{K}_2 \otimes \mathbf{K}_1$ is the Kronecker product of the discretized
kernels $\mathbf{K}_1\in \mathbb{R}^{M_1 \times N_1}$ and $\mathbf{K}_2\in \mathbb{R}^{M_2 \times N_2}$, 
the vector  $\mathbf{s} \in \mathbb{R}^{M}$, $M=M_1 \cdot M_2$, represents
the measured noisy signal, $\mathbf{f} \in   \mathbb{R}^{N}$, $N=N_1 \cdot N_2$, is the vector
reordering of the 2D distribution to be computed and $ \mathbf{e} \in  \mathbb{R}^{M}$ represents the
additive Gaussian noise.

The linear problem \eqref{mod1} is typically ill-conditioned and regularization strategies are necessary to obtain stable discrete distributions. A recent review of regularization techniques used for 2DNMR data inversion can be found in \cite{ComputGeoSci2021}.
\subsection*{The minimization problem}
\Name\ uses a multipenalty approach based on both $L1$ and $L2$ regularization with locally adapted regularization parameters. In this regularization framework, the NMR data inversion problem is reformulated as the unconstrained minimization problem
\begin{equation}
  \min_{\mathbf{f}} \left \{ \|  \mathbf{K} \mathbf{f} - \mathbf{s} \|^2 +\sum_{i=1}^N \lambda_i(\mathbf{L} \mathbf{f})^2_i+ \alpha \|\mathbf{f}\|_1 \right \},
  \label{eq:uno}
\end{equation}
where $\| \cdot \|$ denotes the Euclidean norm.
The first term of the objective function expresses data consistency in the presence of Gaussian noise while the penalty terms take into account two kind of \emph{a priori} information about the underlying distribution: first, the distribution is known to be a smooth function with some Gaussian-like peaks over flat areas and, second, it is known to be sparse. By using the UPEN principle, the values of the regularization parameters $\lambda_i$, $i=1,\ldots,N$ and $\alpha$ can be automatically computed and adapted to the shape of the sought-for distribution \cite{Bortolotti2016}. In our previous work \cite{ComputGeoSci2021}, we have shown that multipenalty regularization \eqref{eq:uno}, involving $L2$ and $L1$ norm, is able to promote distinct features of the computed distribution, since it produces a good trade-off among data fitting error, sparsity and smoothness of the solution.
\subsection*{The minimization method}
The Fast Iterative Shrinkage and Thresholding (FISTA) method with constant backtracking is used to efficiently compute a solution of \eqref{eq:uno}.
We first reformulate problem \eqref{eq:uno} as the sum of two convex functionals:
\begin{equation}
 \min_{\mathbf{f}} \left \{\Psi_1(\mathbf{f})+ \Psi_2(\mathbf{f})\right \}
 \label{eq:due}
\end{equation}
where:
\begin{equation*}
  \Psi_1(\mathbf{f})=  \left \|
\begin{pmatrix}
 \mathbf{K} \\
 \sqrt{\mathbf{\Lambda}} \mathbf{L}
 \end{pmatrix} \mathbf{f} - \begin{pmatrix}
 \mathbf{s} \\
 \mathbf{0}
 \end{pmatrix}
\right \|^2, \;  \mathbf{\Lambda}=\text{diag}(\lambda_i) \;\, \text{and} \;\,
  \Psi_2(\mathbf{f})= \alpha\| \mathbf{f} \|_1 .
\end{equation*}
Given a constant stepsize $\xi$ and a starting point $\mathbf{f}^{(0)}$, the FISTA method for \eqref{eq:due} generates a sequence of iterates as follows
\begin{align}
  \mathbf{f}^{(j)} &= \arg\min_{\mathbf{f}} \left \{  \Psi_2(\mathbf{f}) + \frac{\xi}{2} \left \|  \mathbf{f} - \left (  \mathbf{y}^{(j)} - \frac{1}{\xi} \nabla(\Psi_1(\mathbf{y}^{(j)}))  \right )\right \| \right \}  \label{eq:subproblem}\\
  t_{j+1} &= \frac{1}{2}\left( 1+\sqrt{1+4t_j^2} \right) \\
  \mathbf{y}^{(j+1)} &= \mathbf{f}^{(j)} + \displaystyle{\frac{(t_j-1)}{t_{j+1}}} \left(\mathbf{f}^{(j)}-\mathbf{f}^{(j-1)} \right)
\end{align}
with $\mathbf{y}^{(1)}=\mathbf{f}^{(0)}$ and $t_1=1$. %
Since $\Psi_2(\mathbf{f})= \alpha\| \mathbf{f} \|_1$, the solution of the subproblem \eqref{eq:subproblem} can be computed explicitly, element-wise, by means of the soft thresholding operator:
\begin{equation}
  \mathbf{f}^{(j)}_i = \text{sign}\left (z_i^{(j)}-\frac{\alpha}{\xi} \right) \max\left ( \left |z_i^{(j)} \right |-\frac{\alpha}{\xi} ,0\right ), \ \ i=1, \ldots, N
\label{eq:f_fista}
\end{equation}
where
\begin{equation*}
  \mathbf{z}^{(j)}= \mathbf{y}^{(j)} -  \frac{1}{\xi} \nabla(\Psi_1(\mathbf{y}^{(j)})).
\end{equation*}
Convergence of FISTA has been proved for constant stepsizes $\xi$ equal to a Lipschitz constant of $\nabla(\Psi_1)$ \cite{Beck_Teb_2013}. In \Name, we set
\begin{equation}
   \xi=\left (\sigma_1^{(1)} \sigma_1^{(2)} \right )^2 + 64  \max_i | \lambda_i |
   \label{eq:LL}
\end{equation}
where $\sigma_1^{(1)}$ and $ \sigma_1^{(2)} $ are the maximum singular values of the matrices $\mathbf{K}_1$ and $\mathbf{K}_2$, respectively. The choice \eqref{eq:LL} for the stepsize $\xi$ can be proved to satisfy \cite{ComputGeoSci2021}
\begin{equation}\label{bound}
  \lambda_{\max} ( \mathbf{K}^T  \mathbf{K} + \mathbf{L}^T \mathbf{\Lambda} \mathbf{L}) \leq \xi
\end{equation}
where $\lambda_{\max}(\cdot)$ denotes the maximum eigenvalue of a matrix. The lower bound \eqref{bound} ensures FISTA convergence \cite{Beck_Teb_2013}.
\subsection*{Spatially adapted regularization parameters}
The choice of the regularization parameters $\lambda_i$, $i=1,\ldots,N$, and $\alpha$ is crucial to obtain a meaningful distribution. In \cite{Bortolotti2016,ComputGeoSci2021}, we developed an automatic selection scheme for spatially adapted regularization parameters by using the UPEN principle. Numerical examples in our works demonstrate the effectiveness of this selection rule.

Given an approximated distribution $\mathbf{f}$, our automatic selection rule for the regularization parameters can be described as follows:
\begin{align}
    \alpha &= \displaystyle{\frac{\|\mathbf{K} \mathbf{f} - \mathbf{s}\|^2}{(N+1)  \| \mathbf{f} \|_1}}; \label{eq:alpha} \\
    \lambda_i &= \frac{\|\mathbf{K} \mathbf{f} - \mathbf{s} \|^2}%
    {(N+1)\left ( \beta_0 +\beta_p\underset{\substack{\mu \in I_i}}\max \, \{\mathbf{\text{vec} \big( \| \nabla \mathbf{F}\|\big)}_{\mu}^2\} %
    + \beta_c \underset{\substack{\mu \in I_i }} \max \, \{\mathbf{(\mathbf{L}\mathbf{f}})_{\mu}^2\}\right )},  \quad i=1,\ldots,N; \label{eq:lambda}
\end{align}
where $\text{vec}(\cdot)$ denotes the operator that stacks a matrix column-wise to produce a column vector and $\mathbf{F}$ is the 2D distribution map corresponding to $\mathbf{f}$ (i.e., $\mathbf{f}=\text{vec}(\mathbf{F})$).
The indices subsets $I_i$, $i=1,\ldots,N$, are related to the neighborhood of the point $i$
and the $\beta$'s are positive parameters whose optimum values can change with the nature of the measured sample. Please refer to \cite{Bortolotti2016}, for a detailed description of the properties of the $\beta$
parameters.
%
%
\subsection*{The implemented algorithm}
The method implemented in \Name\ computes the solution of a weighted inversion problem based on  \eqref{eq:uno}, i.e.:
 \begin{equation}
  \min_{\mathbf{f}} \left \{ \|  \mathbf{K} \mathbf{f} - \mathbf{s} \|^2 + \omega_1 \sum_{i=1}^N \lambda_i(\mathbf{L} \mathbf{f})^2_i+ \omega_2 \left (\alpha \|\mathbf{f}\|_1 \right ) \right \}, \ \ \omega_1,\omega_2 \in [0,1]
  \label{eq:tre}
\end{equation}
where the default values of the weights are $\omega_1=\omega_2=1$, as in \eqref{eq:uno}.
Depending on the data and problem type, it is possible to introduce further model flexibility by setting $\omega_i$ s.t. $\omega_1+\omega_2=1$. 
This feature is implemented by setting a specific flag discussed in the subsequent section.\\
Problem \eqref{eq:tre} is solved by the FISTA algorithm (equations: \eqref{eq:due}-\eqref{eq:f_fista}), using the  rules \eqref{eq:alpha}-\eqref{eq:lambda} for the automatic computation of the spatially adapted regularization parameters. We report the steps of our method in Algorithm \ref{alg:L1LL2}.
\begin{algorithm}[h!]
\caption{\label{alg:L1LL2}}
\vskip 1mm
\begin{spacing}{1.15}
\begin{algorithmic}[1]
\STATE Choose $\tau\in(0,1)$, $\beta_0,\beta_p,\beta_c>0$ and compute $\sigma_1^{(1)}$ and $\sigma_1^{(2)}$.
\STATE \emph{\textbf{Initialization}}\\
Set $k=0$ and choose a starting guess $\mathbf{f}^{(0)}$.
\REPEAT
  \STATE \emph{\textbf{Regularization parameters update}}\\
  Set
  \begin{align*}
    \alpha^{(k)} &= \displaystyle{\frac{\|\mathbf{K} \mathbf{f}^{(k)} - \mathbf{s}\|^2}{(N+1)  \| \mathbf{f}^{(k)} \|_1}}; \\
    \lambda_i^{(k)} &= \frac{\|\mathbf{K} \mathbf{f}^{(k)} - \mathbf{s} \|^2}%
    {(N+1)\left ( \beta_0 +\beta_p\underset{\substack{\mu \in I_i}}\max \, \{\mathbf{\text{vec} \big( \| \nabla \mathbf{F}^{(k)}\|\big)}_{\mu}^2\} %
    + \beta_c \underset{\substack{\mu \in I_i }} \max \, \{\mathbf{(\mathbf{L}\mathbf{f}^{(k)}})_{\mu}^2\}\right )},  \quad i=1,\ldots,N;
  \end{align*}
  \STATE \emph{\textbf{Solution update}}\\
  Compute
  \begin{equation*}
    \mathbf{f}^{(k+1)} = \text{argmin}_{\mathbf{f}} \left\{%
    \left \| \begin{pmatrix}
        \mathbf{K} \\
        \sqrt{\mathbf{\omega_1\Lambda}^{(k)}} \mathbf{L}
        \end{pmatrix} \mathbf{f} - \begin{pmatrix} \mathbf{s} \\ \mathbf{0} \end{pmatrix} \right\|^2
    +\omega_2\alpha^{(k)} \| \mathbf{f} \|_1
    \right\}
  \end{equation*}
  by the FISTA method with starting guess $\mathbf{f}^{(k)}$ and constant stepsize %
  $\xi^{(k)}=\big(\sigma_1^{(1)} \sigma_1^{(2)} \big )^2 + 64  \max_i | \lambda_i^{(k)} |$
\UNTIL{$\|\mathbf{f} ^{(k+1)} - \mathbf{f} ^{(k)} \| \leq \tau \|  \mathbf{f} ^{(k)} \|$}
\end{algorithmic}
\end{spacing}
\end{algorithm}
Concerning the  starting guess $\mathbf{f}^{(0)}$ we apply a few iterations of the Gradient Projection (GP) method to the non-negatively constrained least squares problem, i.e.
\begin{equation*}
  \mathbf{f}^{(0)}= \arg \min_{\mathbf{f} \geq 0}  \|\mathbf{K} \mathbf{f} - \mathbf{s} \|^2 .
\end{equation*}

\section*{Software structure and design \label{toolbox}}
\begin{figure}[h!]
    \centering
    \includegraphics[scale=0.3]{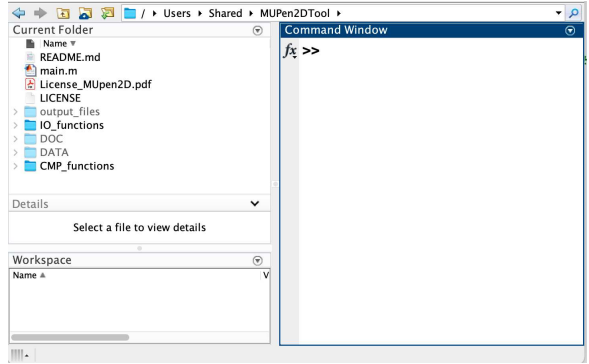}
    \caption{MATLAB {\em current folder} to run \Name.}
    \label{fig:MatPath}
\end{figure}

A MATLAB toolbox is a collection of user-written MATLAB files, with functions and/or classes, aimed at addressing a specific topic.
\Name\ is a collection of functions related to different aspects of the inversion of 2DNMR data.
The package, available at \url{https://doi.org/10.6084/m9.figshare.18197399}, contains all MATLAB scripts and functions to run \Name\  from the MATLAB environment. Alternatively, it is possible to test \Name\ using  the MATLAB app {\tt GUI\_MUPEN2D}  with the dedicated (stand alone compiled) GUI application available from \url{https://site.unibo.it/softwaredicam/en/mupen2d}. 
After downloading  {\tt MUPen2DTool.zip}, extract the content to the folder \Name, set the MATLAB {\em current folder} as reported in figure \ref{fig:MatPath}, and
%
run the script {\tt main.m} in the MATLAB command window. 

The  implemented software functions  are contained  into the following sub-folders:
\begin{itemize}
    \item {\tt IO\_functions}, related to {\em Initialization, data load, output data and figures}, contains the following functions:
     {\tt SetInputFile.m}, {\tt SetPar.m}, {\tt LoadData.m}, {\tt LoadFlags.m}, {\tt grafico\_1D.m}, {\tt grafico\_2D.m}, {\tt grafico\_3D.m}.
    \item {\tt CMP\_functions}, implementing the inversion algorithm and post-processing, contains:  {\tt MUpen2D.m},
    {\tt T1\_T2\_Kernel.m}, {\tt T2\_T2\_Kernel.m}, {\tt D\_T2\_Kernel.m}, {\tt get\_diff.m}, {\tt get\_l.m}, {\tt Residual\_Analysis.m}.
    \item {\tt DATA}, containing the inversion data sub-folders, is  described in details in the sequel.
    \item {\tt DOC}, contains the user manual.
     \item {\tt output\_files} is generated by {\tt main} and contains data and inversion information, refer to the software Manual for more details.
\end{itemize}
\paragraph{ Description of {\tt Data} folder.}
The software has included  the following data folders for testing:
\begin{itemize}
    \item {\tt T1-T2\_synth\_2pks} relative to a synthetic $T_1-T_2$ data numerically generated;
    \item {\tt T1-T2\_EDTA\_Triple\_IRCPMG} relative to $T_1-T_2$ data from a synthetic EDTA physical sample; 
    \item {\tt D-T2} relative to a $D-T_2$ data from a fresh cement sample.
\end{itemize}
Each data folder contains six files: three data files,
\begin{itemize}
\item a column ASCII data file with the list of the $M$ time values used to acquire the relaxation curve on the first dimension, 
\item a column ASCII data file with the list of the $N$ time values used to acquire the relaxation curve on second dimension, 
\item a 2D $M   \times N$  matrix data file with the acquired signal, 
\end{itemize} 
and three parameter files: {\tt Filepar.par}, {\tt FileFlag.par}, {\tt FileSetInput.par} containing the keywords that define the type of experiment to be processed and to modify the functionality of \Name. 
The detailed description of each keyword can be found in the user manual, here we only outline the main structure. \\
The file {\tt FileSetInput.par} contains the name of the 2D data file {\tt Filenamedata}, the names of the files containing the vectors of times in the  first  ({\tt filenameTimeX}) and second ({\tt filenameTimeY}) dimensions and number the relaxation time bins along the first ({\tt nx}) and second ({\tt ny}) dimension. 

In {\tt FileFlag.par} we select the type of relaxation data ({\tt FL\_typeKernel}), the  left and right extreme limits of the relaxation times ({\tt FL\_InversionTimeLimits}).

Finally {\tt FilePar.par} contains the parameters of the inversion algorithm, such as tolerances, maximum number of iterations allowed, the weights in $L2$ penalty function \eqref{eq:lambda} ($\beta_0, \beta_p, \beta_c$) and the value {\tt par.fista.weight} to set the  weights $\omega_1, \omega_2$ as follows:
$$
\begin{array}{l}
 {\tt if \  par.fista.weight} \in [0,1], \ \left \{
 \begin{array}{l} 
 \omega_1=1-{\tt par.fista.weight} \\
 \omega_2={\tt par.fista.weight} 
 \end{array} \right . \\
  {\tt elseif  \ par.fista.weight}\notin [0,1], \    \omega_1=\omega_2=1
  \end{array}
$$
Usually there is no need to set ${\tt par.fista.weight} \in [0,1]$ however, depending on the acquired data, the results might be improved. 

\section*{Examples \label{valid}}
In this section we present a few examples of the results computed by \Name\ on the dataset included in the software package, with the purpose to help users to  adapt their own data.
The included data folder contains two examples of $T_1-T_2$ data and one example of $D-T_2$ data. The first $T_1-T_2$ example is relative to synthetic data numerically generated from a known map of relaxation times. The second $T_1-T_2$ example contains an acquisition performed on a synthetic physical sample (a self-made sample). The $D-T_2$ data are relative to a fresh Waite Portland Cement (WPC) sample.

In the sequel, capital letters denote matrices and the corresponding small letters denote the vector obtained by matrix reordering. 

The reported results have been obtained by using a PC equipped with a $2.9$ GHz Intel microprocessor i7 quad-core and 16 GB RAM.

\subsection*{$T_1-T_2$  synthetic numerical data \label{T1T2Sy}}
This example consists of a synthetic data generated by the source code {\tt main\_make\_synt.m} contained in the folder {\tt T1-T2\_synth\_2pks}.
To run this test we must select the folder {\tt T1-T2\_synth\_2pks} either launching the {\tt main.m} script in the MATLAB environment or directly the GUI application.

\begin{figure}
    \centering
    \includegraphics[scale=0.3]{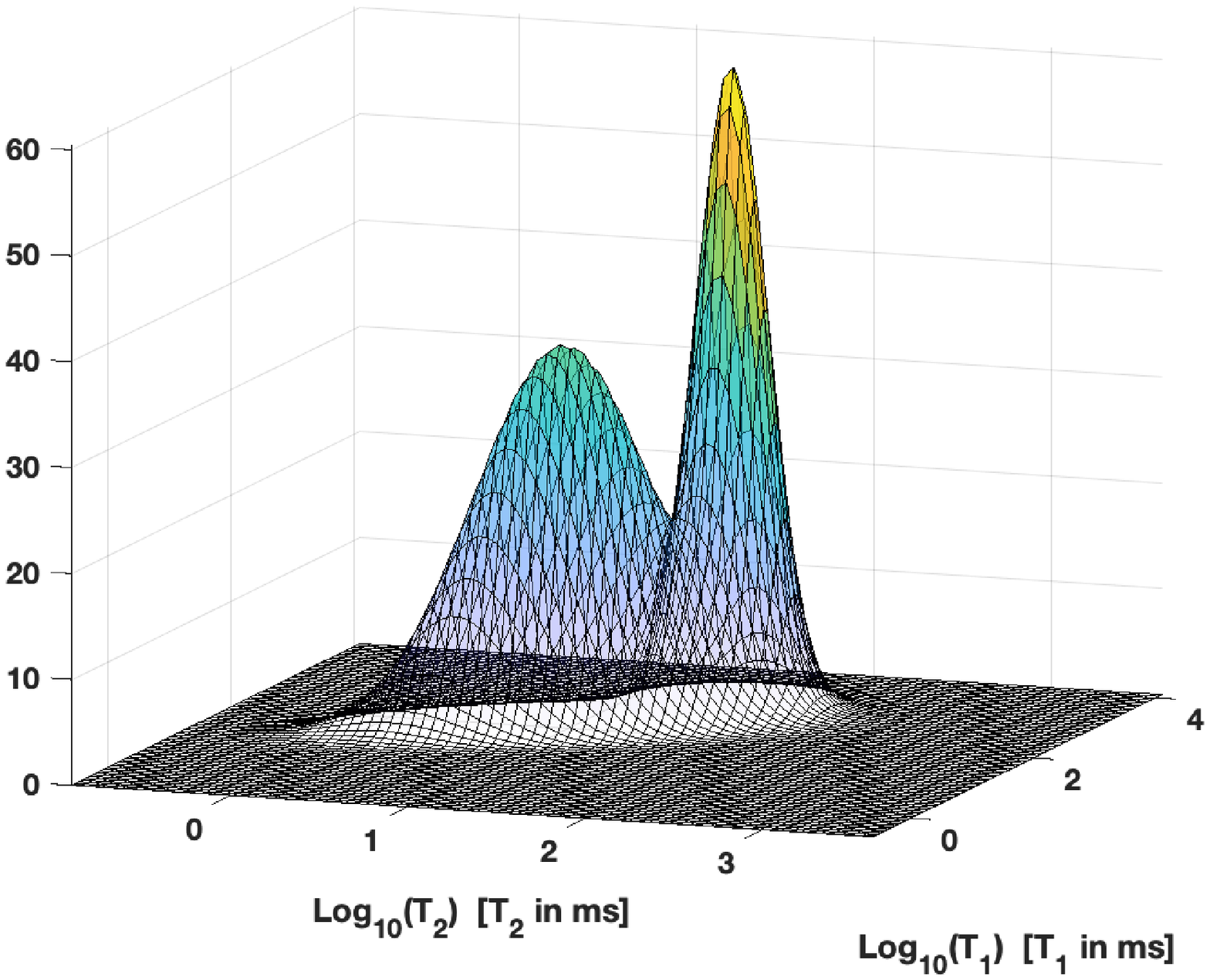}
    \includegraphics[scale=0.3]{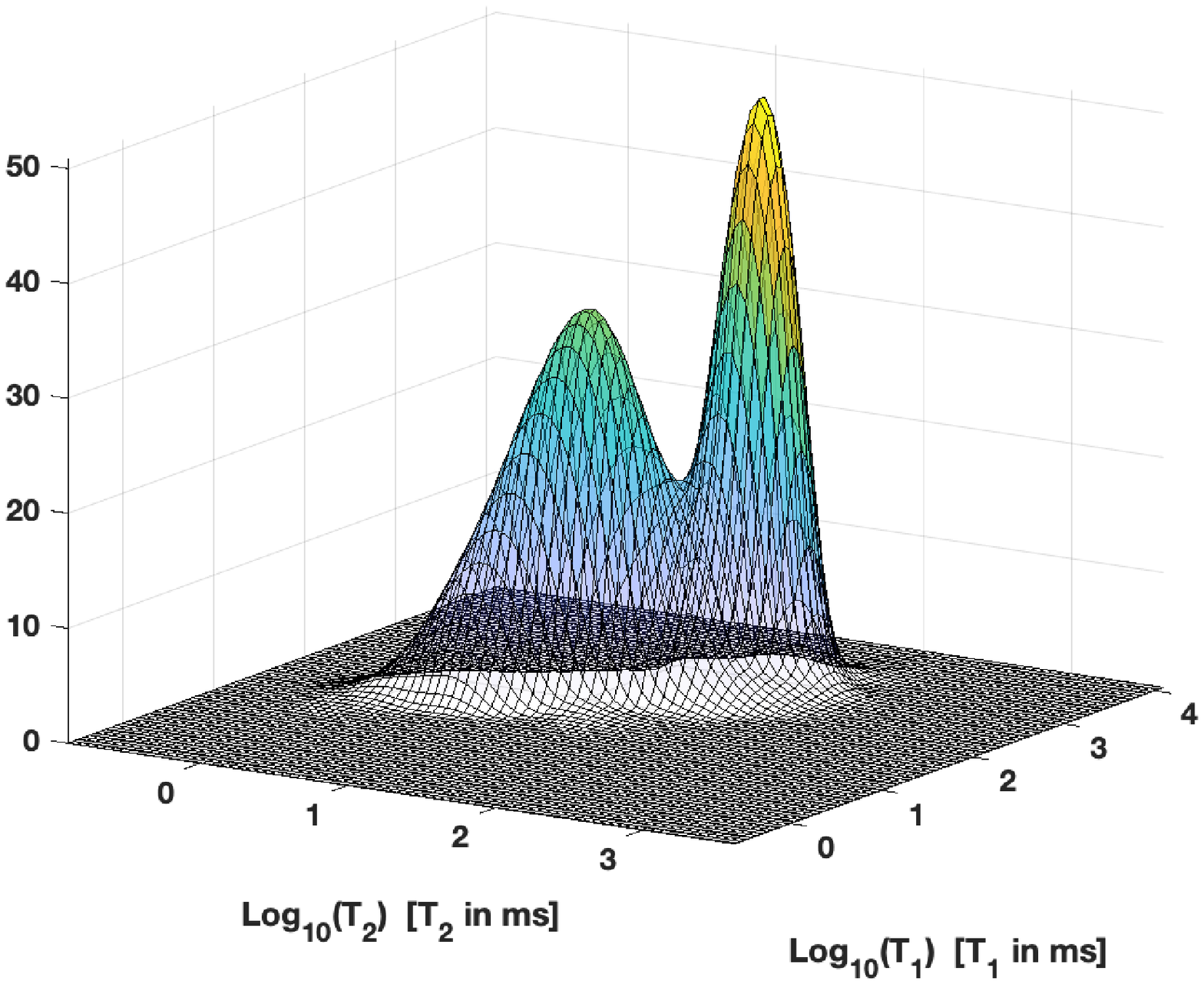}
     \\
    (a) \hspace{2cm} (b)
    \caption{Relaxation map: (a) reference, (b) computed by \Name.}
    \label{fig:SY_2}
\end{figure}

The reference relaxation map $F(T_1,T_2)$, represented in Figure \ref{fig:SY_2} (a), has size $80\times 80$ and presents two peaks at positions ($T_1=815.0 \ ms$, $T_2=4.533 \ ms$) and ($T_1 = 119.5 \ ms$, $T_2=8.561 \ ms$). The reference map is applied to \eqref{eq:modelT1T2} with the IR kernel  to obtain the synthetic relaxation data $S(t_1,t_2)$ corresponding to an IRCPMG experiment. The number of IR inversion times  is $M_1= 128$ while the CPMG sequence has $M_2 = 2048$ echoes.
Normal Gaussian random noise  $\mathbf{e}\in\mathbb{R}^{128\times2048}$ of level $\delta\equiv\|\mathbf{e}\|=10^{-2}$ is added.\\
 
The computed relaxation map displayed in  Figure \ref{fig:SY_2} (b) shows an accurate representation of the reference map.
The detailed discussion of  this test in terms of errors and comparison with different solution strategies can be found in \cite{ComputGeoSci2021}. 

\begin{figure}
    \centering
    \includegraphics[scale=0.7]{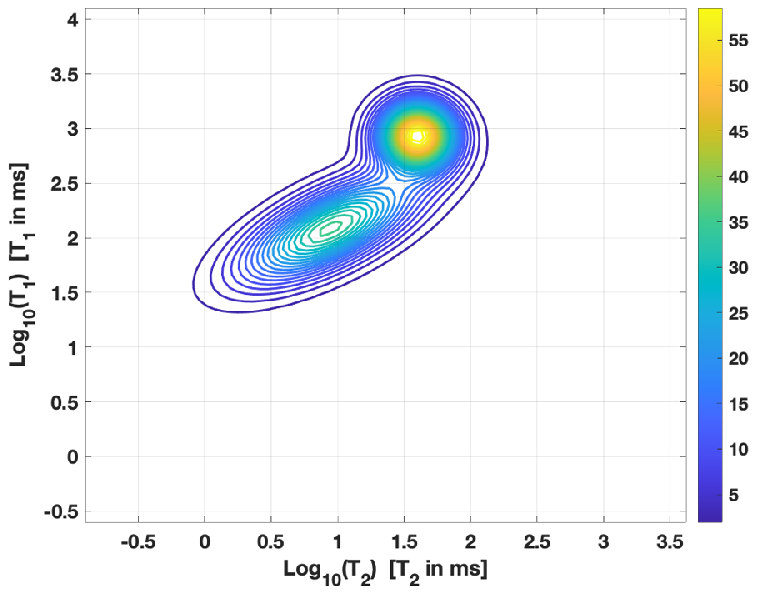}
    \includegraphics[scale=0.3]{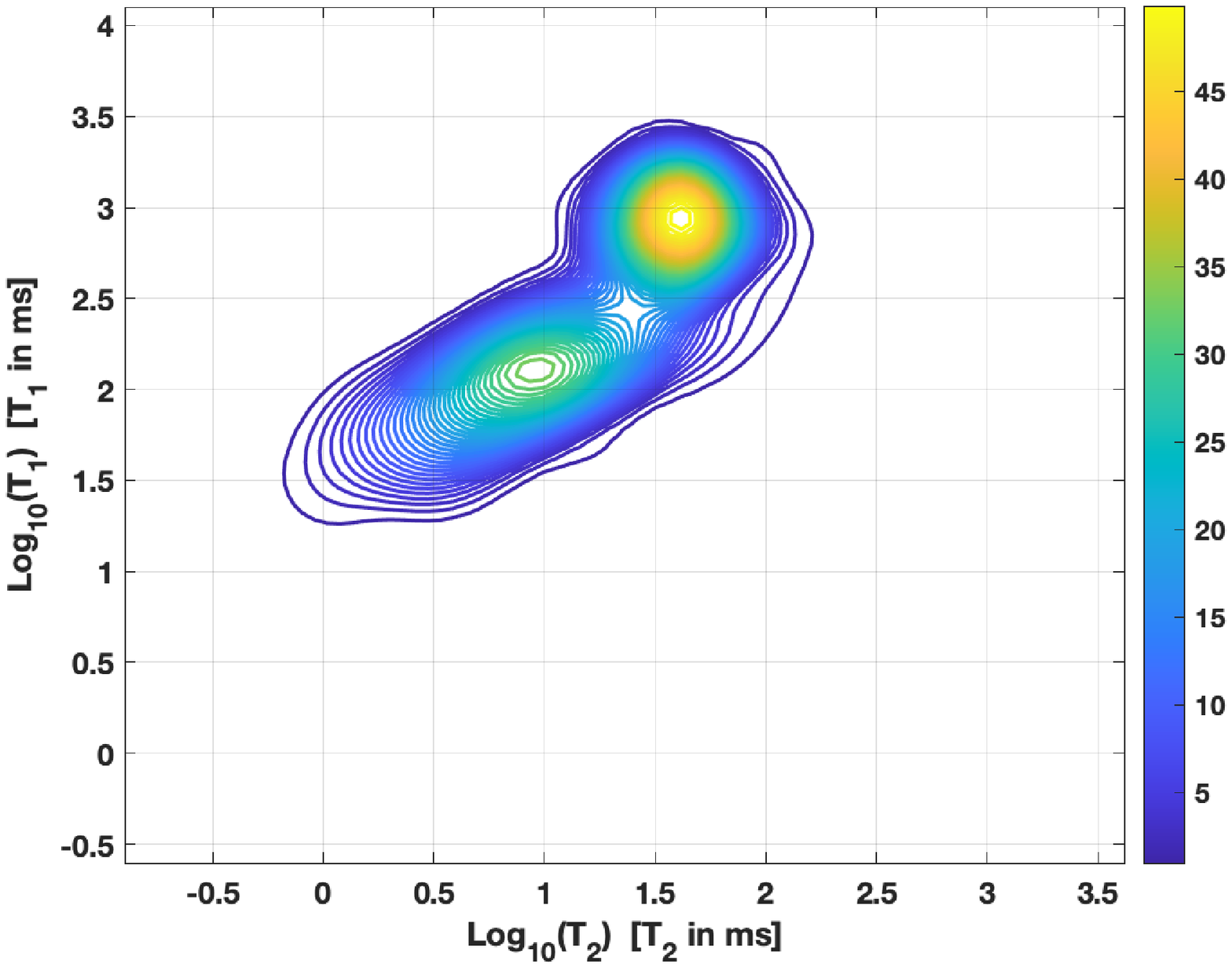} \\
    (a) \hspace{2cm} (b)
    \caption{Contour maps: (a) reference, \ (b) computed by \Name.}
    \label{fig:2pks_1}
\end{figure}

\begin{figure}
    \centering
    \includegraphics[scale=0.3]{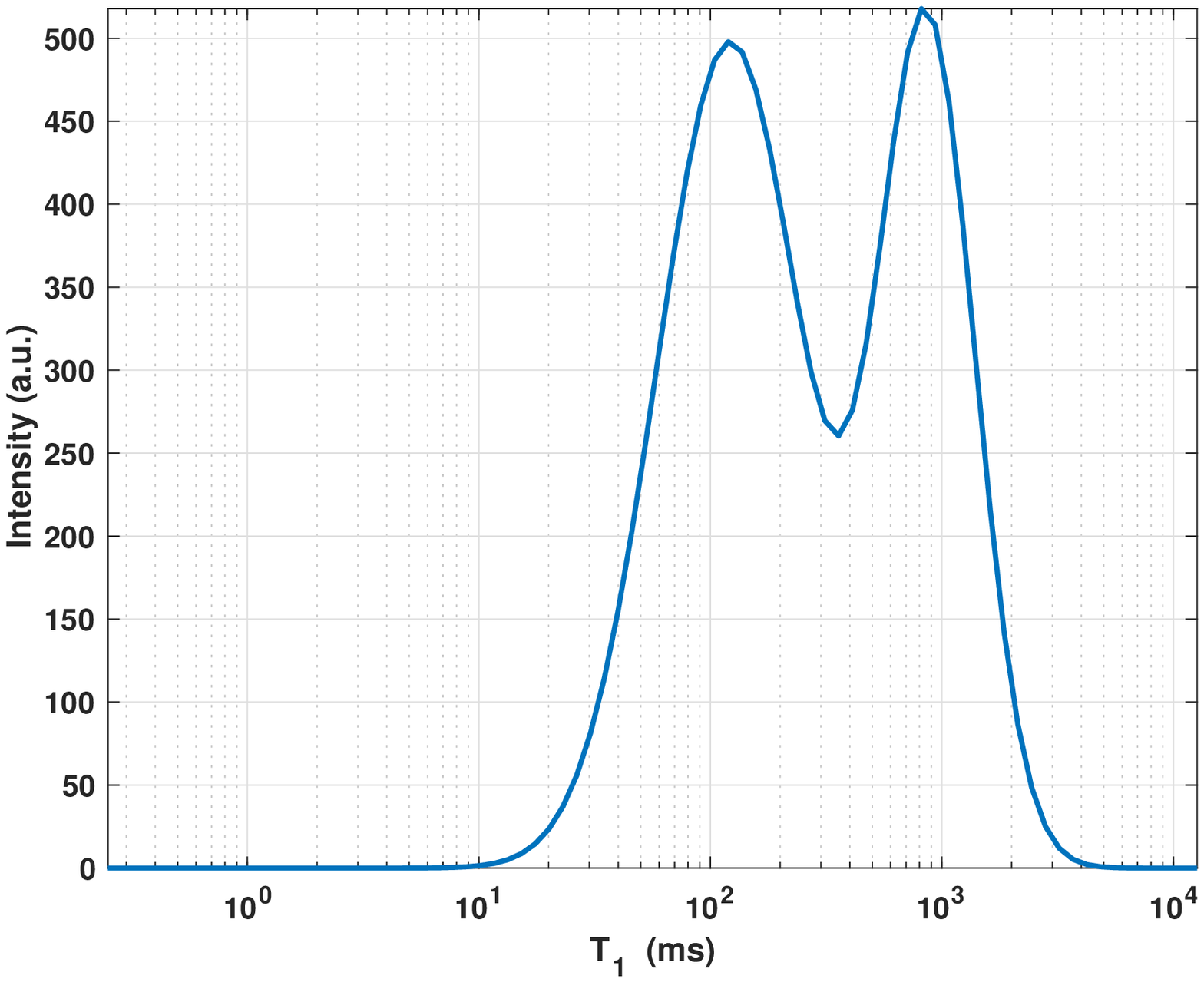}
    \includegraphics[scale=0.3]{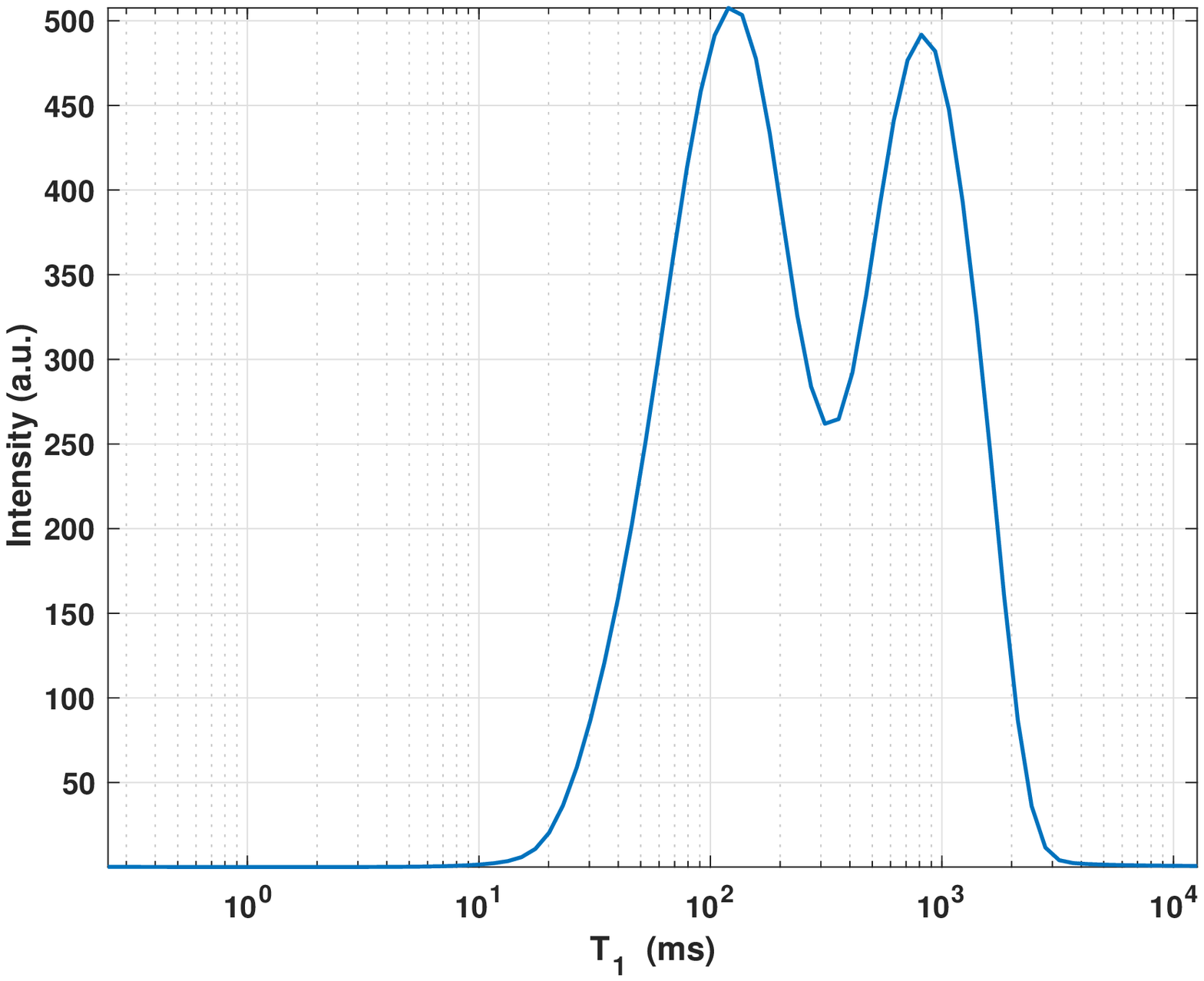} \\
    (a) \hspace{2cm} (b)
    \caption{$T_1$ one dimension projection: (a) reference, \ (b) computed by \Name.}
    \label{fig:2pks_T1}
\end{figure}

\begin{figure}
    \centering
    \includegraphics[scale=0.3]{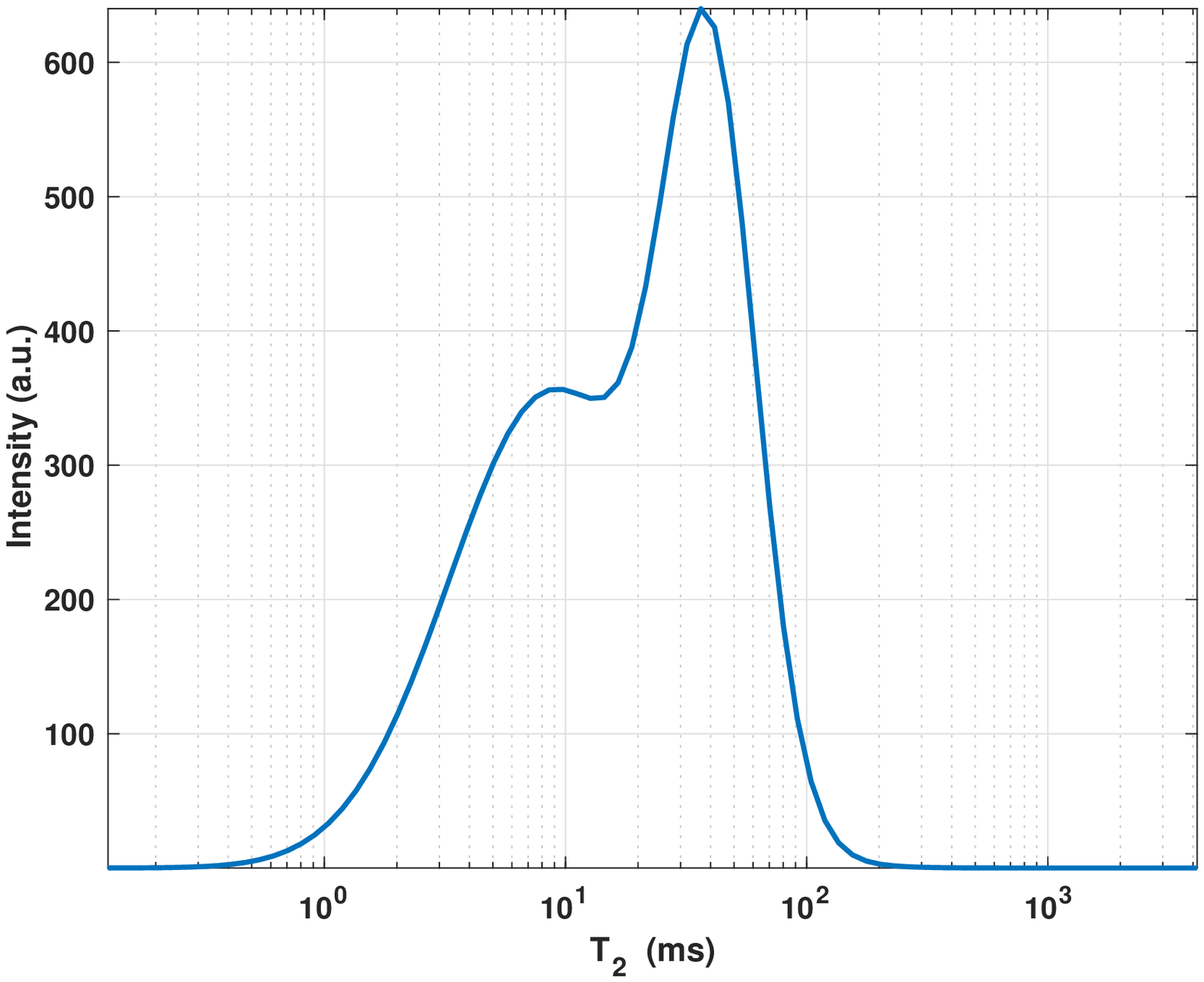} 
    \includegraphics[scale=0.3]{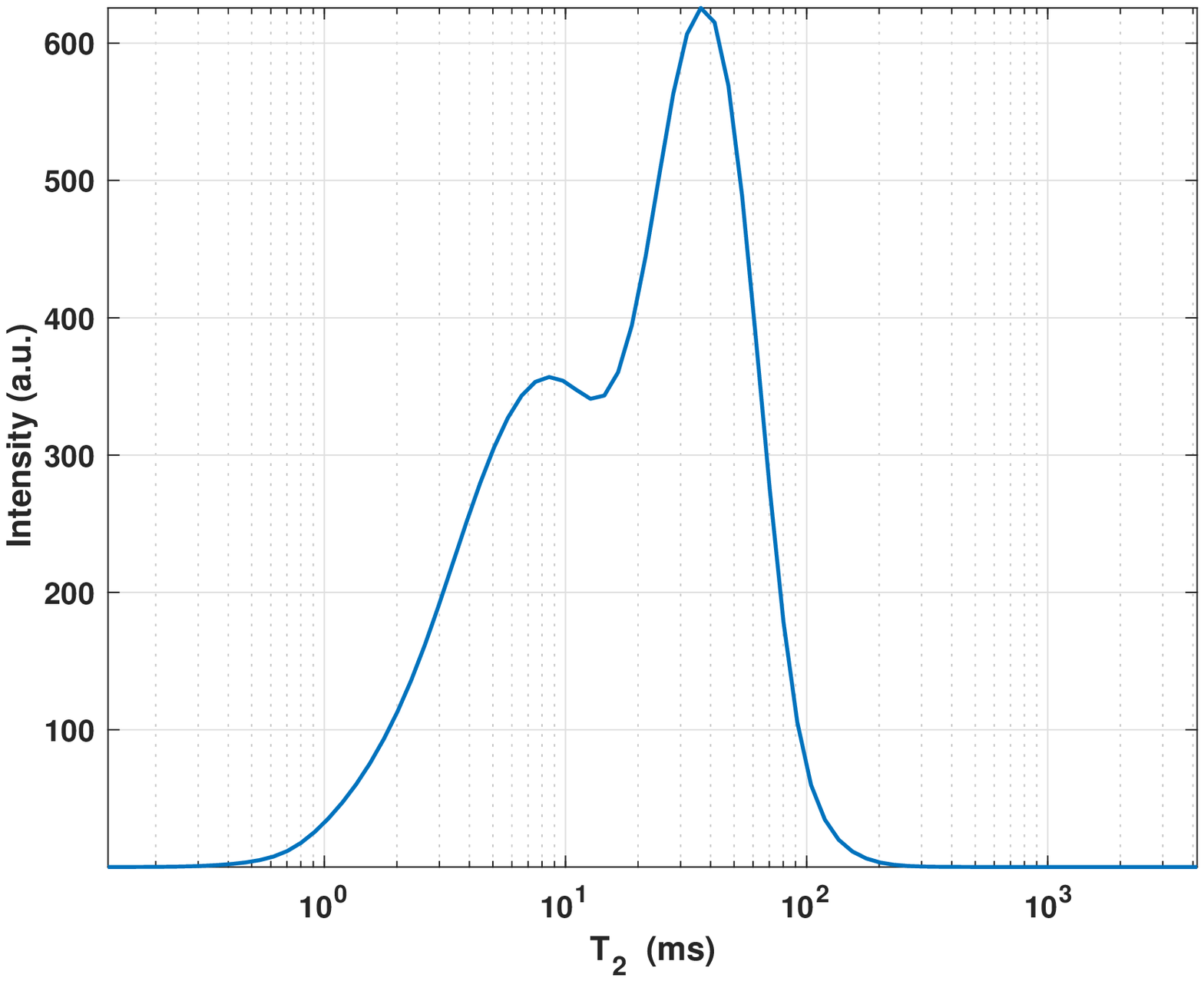}
     \\
    (a) \hspace{2cm} (b)
    \caption{$T_2$ one dimension projection: (a) reference, \ (b) computed by \Name.}
    \label{fig:2pks_T2}
\end{figure}

\begin{figure}
    \centering
   \includegraphics[scale=0.3]{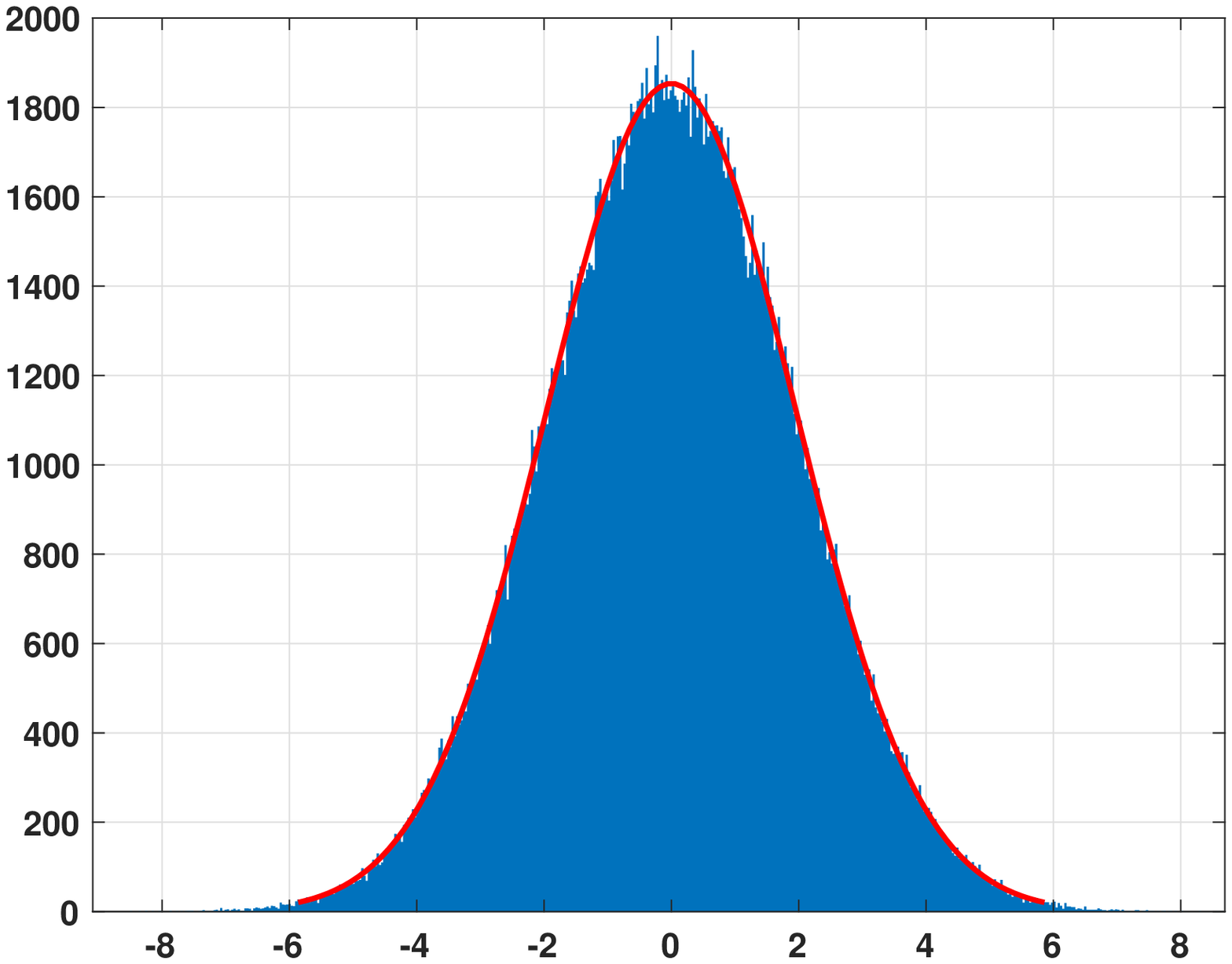}    
    \includegraphics[scale=0.3]{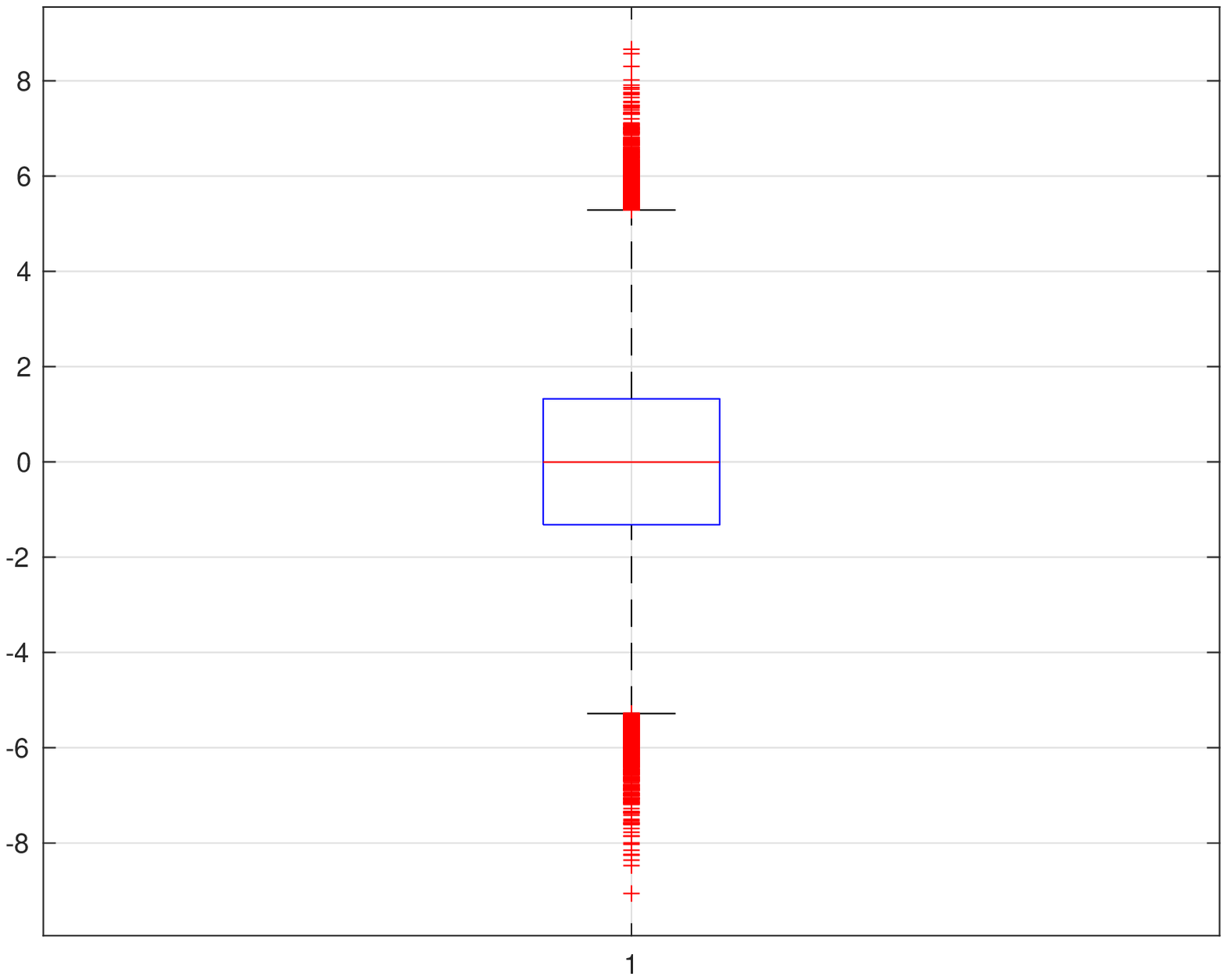} \\
(a) \hspace{2cm} (b)
    \caption{Post-Processing. (a) Histogram of Residual values (blue bars) and Fitted Normal distribution  (red line) (b) Box Plot:
    red horizontal line represents the median value ($-4.0074 \ 10^{-3})$;
    bottom and top edges of the box contain the $25$th  and $75$th  percentiles;
    red   '+' marker represents the outliers.}
    \label{fig:2pks_hist}
\end{figure}

Figure \ref{fig:2pks_1} shows the contour map computed by \Name\ and the contour reference map from which we can appreciate the accurate localization of the peaks.
The $T_1$, $T_2$ projections are reported in figures \ref{fig:2pks_T1} and \ref{fig:2pks_T2} for both the \Name\ computed map and the reference map.
%
The post-processing step computes some interesting information about the residual $\mathbf{R}=\mathbf{s}- \mathbf{K} \mathbf{f}_\text{computed}$.
The histogram of the residual $\mathbf{R}$, in figure \ref{fig:2pks_hist}(a), shows the good agreement of the residual values (blue bars) to the normal distribution (red line) with variance $\sigma=1.9531$ and mean $\mu=3.9436 \ 10^{-5}$.
Moreover, the box-plot in figure \ref{fig:2pks_hist}(b) represents the box having  the bottom edge given by the $25$th percentile ($-1.3185$) and the top edge given by the $75$th  percentile ($ 1.3233$), containing $260444$ out of $262144$ points ($99.35  \%$). The median value ( $-4.0074 \ 10^{-3}$) is represented as an horizontal red line. The $1700$ outliers are represented by red '+' markers.
Information about the lack of symmetry (skewness) and weight of the distribution tails (kurtosis) is contained in the structure {\tt out\_data}, i.e.:
$$\hbox{{\tt out\_data.skewness}} = -5.2766 \ 10^{-4}, \ \ \ \hbox{{\tt out\_data.kurtosis}}= 2.9833.$$
Following \cite{hair2010},  $\mathbf{R}$ is considered to be normal since  skewness is between $-2$ and $+2$ and kurtosis is between $-7$ and $+7$, demonstrating the goodness of the inversion performed.

The file {\tt Parameters.txt} in the {\tt output} subfolder reports the following information about input tolerances, data size, algorithm iterations and computation time.
\begin{verbatim}
---------------------------------------------------------------------
MUpen2D Input Parameters 
 MUpen2D tol = 1.000000E-04,
 Projected Gradient Tol = 1.000000E-04 
 SVD Threshold = 1E-16 
 Data size = 80 x 80  
---------------------------------------------------------------------
Number of Inversion channels:  horizontal 80, vertical  80 
Final Relative Residual Norm = 2.5421E-03 
Total MUpen2D Iterations = 5
Total FISTA Iterations = 76793 
Computation Time = 26.65099 s
---------------------------------------------------------------------
\end{verbatim}

\subsection*{$T_1-T_2$ dataset: Triple EDTA  \label{T1T2EDTA}}
This test can be performed by selecting the folder “EDTA Triple IRCPMG” with the default parameters set in the files {\tt FileFlag.par}, {\tt FileFlag.par}, {\tt FileSetInput.par}. The acquired raw data are relative to a sample composed of three different glass tubes filled with mixtures of distilled water and CuEDTA at various concentrations. The copper-based additive shortens water relaxation times $T_1$ and $T_2$ making the whole sample characterized by three distinctive relaxation components. The IR-CPMG experiment was performed by a KEAII console with the following parameters: for IR, we used 128 log-spaced inversion times in the range 0.5 $ms$ – 3000 $ms$, and for CPMG we set the number of echoes NE = 8000 with echo time TE = 200 $\mu$s. Size of the computed map was 64 × 64.
Moreover, to evaluate the quality of the results of 2D inversion, one-dimensional IR and CPMG relaxation curves were independently acquired by a Stelar relaxometer and analyzed by {\tt Upenwin} software. 

\begin{figure}[h!]
    \centering
    \includegraphics[scale=0.3]{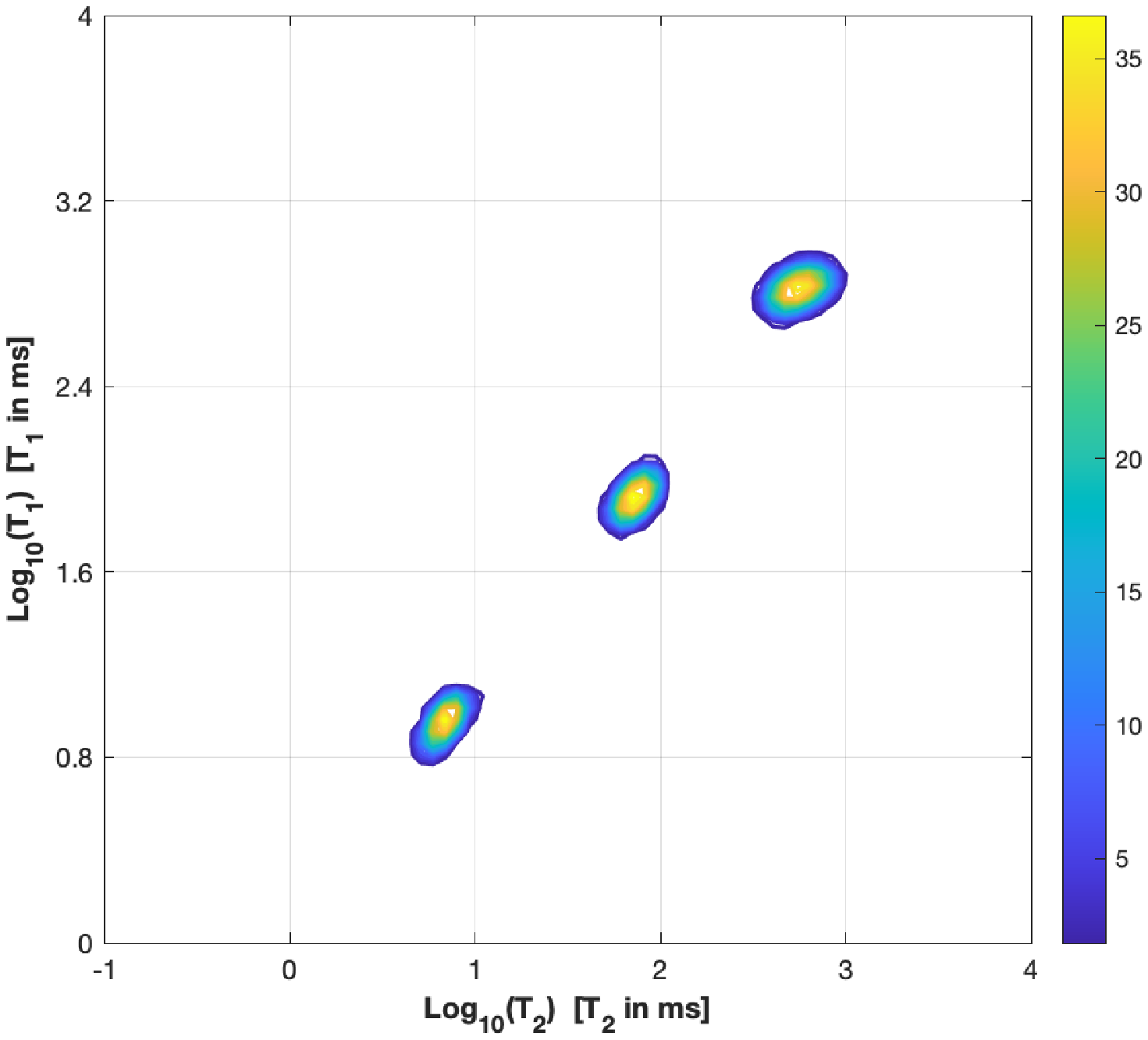}
    \includegraphics[scale=0.3]{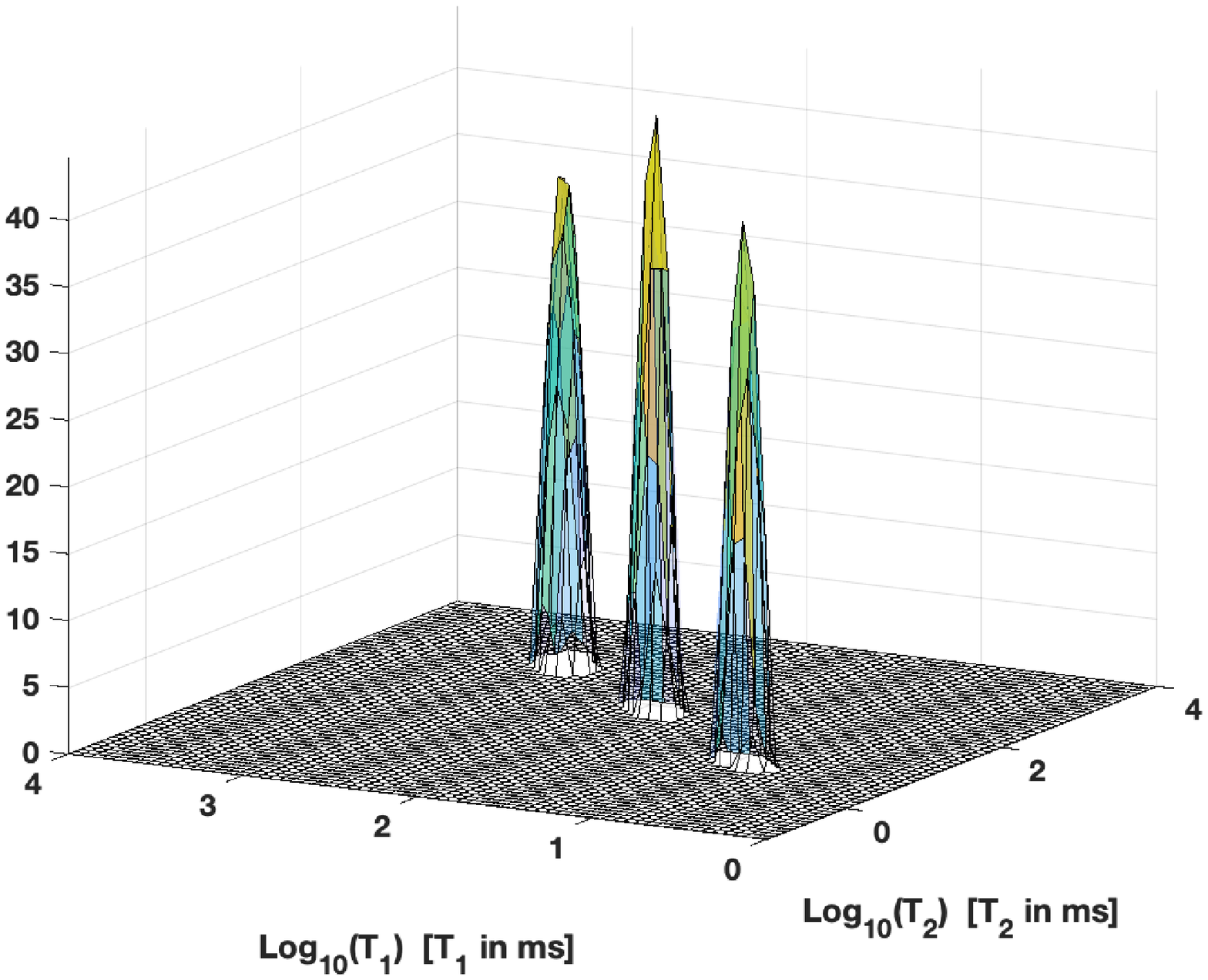} 
  \\
    (a) \hspace{2cm} (b)
    \caption{Contour map (a) and relaxation map (b) computed by \Name.}
    \label{fig:triple_1}
\end{figure}
Figure \ref{fig:triple_1} shows the contour map and the relaxation map computed by \Name. Three well separated components (peaks) were calculated with average relaxation times ($T_1$,$T_2$): C1 = (8.16,6.31) $ms$, C2 = (81.4,71.6) $ms$, C3 = (670,548) $ms$. 
\begin{figure}[h!]
    \centering
    \includegraphics[scale=0.7]{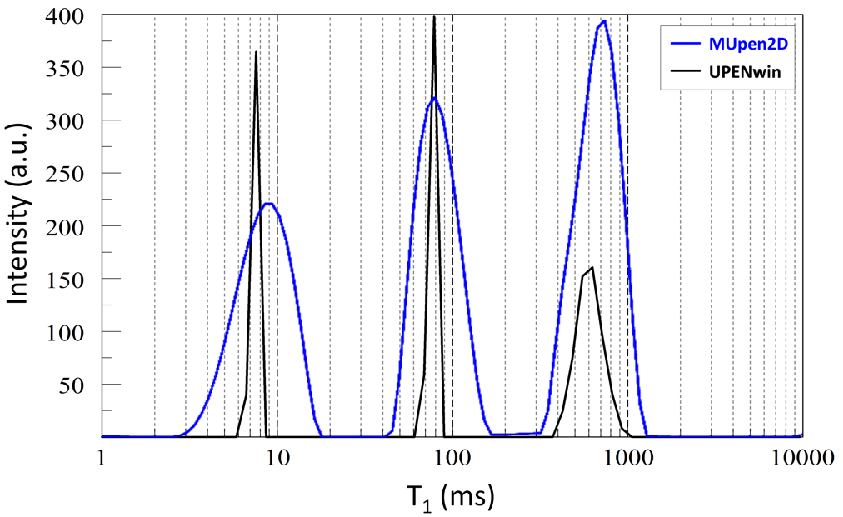}
    \includegraphics[scale=0.7]{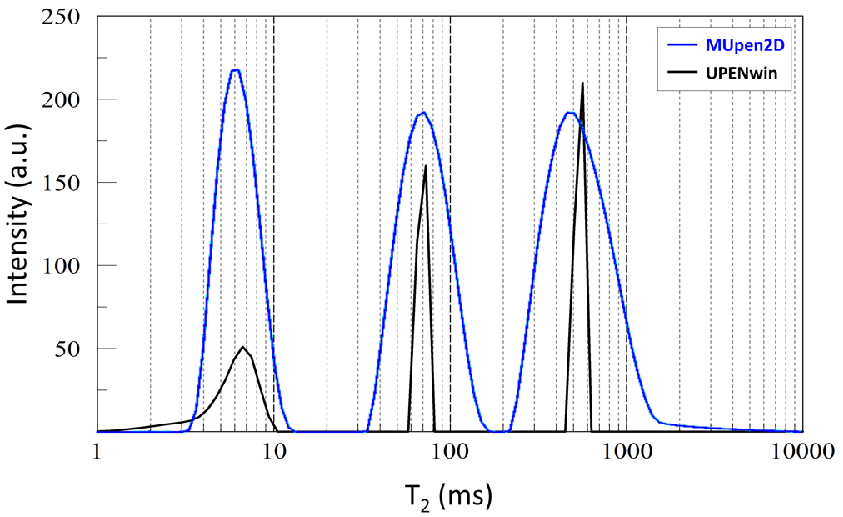} \\
    (a) \hspace{2cm} (b)
    \caption{ $T_1$ one dimension projection (a) and  $T_2$ one dimension projection (b) with overlapped the corresponding 1D distribution computed with {\tt Upenwin}.}
    \label{fig:triple_T12}
\end{figure}
To further quantify the performance, one-dimensional inversions by UPENwin software of independent IR and CPMG acquisitions were discussed. The analysis compared the weighted geometric mean values ($T_{gm}$) of the three $T_1-T_2$ peaks position and their relative areas (\% to the total signal) as computed by \Name\ and {\tt Upenwin} (see Table \ref{tab:tableX} and figure \ref{fig:triple_T12}). The relative differences among computed values are very small for all the considered parameters, showing a good agreement between the two algorithms and confirming the accuracy of the \Name\ computation.
\begin{table}
\centering
\begin{tabular}{lllllll}
\vcell{}         & \vcell{}         & \multicolumn{2}{l}{\vcell{{\tt UPENwin}}} & \vcell{}         & \multicolumn{2}{l}{\vcell{{\tt MUpen2D}}}  \\[-\rowheight]
\printcellbottom & \printcellmiddle & \multicolumn{2}{l}{\printcellmiddle}                       & \printcellmiddle & \multicolumn{2}{l}{\printcellmiddle}                  \\
                 & Peak             & $T_{gm}$ ($ms$) & Pct \%                                          &                  & $T_{gm}$ ($ms$) & Pct \%                                     \\
                 & 1°               & 7.49     & 28.5                                            &                  & 8.16     & 28.3                                       \\
$T1$               & 2°               & 77.3     & 32.2                                            &                  & 81.4     & 31.7                                       \\
                 & 3°               & 603      & 39.3                                            &                  & 670      & 40                                         \\
\multicolumn{7}{l}{}                                                                                                                                                        \\
                 & 1°               & 5.59     & 31.8                                            &                  & 6.31     & 28.3                                       \\
$T2$               & 2°               & 69.4     & 30.8                                            &                  & 71.6     & 31.6                                       \\
                 & 3°               & 541      & 37.4                                            &                  & 548      & 40.1  
\end{tabular}
\caption{\label{tab:tableX} Comparison among results obtained with {\tt Upenwin} and  \Name.}
\end{table}

\subsection*{$D-T_2$ dataset: cement sample \label{DT2}}
The acquired raw data are relative to a fresh WPC sample measured with $D-T2$ maps are done with Stimulated Echo Diffusion Editing sequence SSE-CPMG with the use of a NMR Mouse PM10 set-up (Magritek, NZ) with the field gradient of 14 $T/m$, size of the sensitive volume x,y,z [15x15x(0.1-0.3)] $mm^3$, magnetic field in the sensitive volume 0.327 T driven with a KEAII console (Magritek, NZ) and using the following acquisition parameters: CPMG TE= 400 $\mu$s, Number of echoes 200, number of scans 200, repetition delay 1.5 $s$.

\begin{figure}[h!]
    \centering
    \includegraphics[scale=0.3]{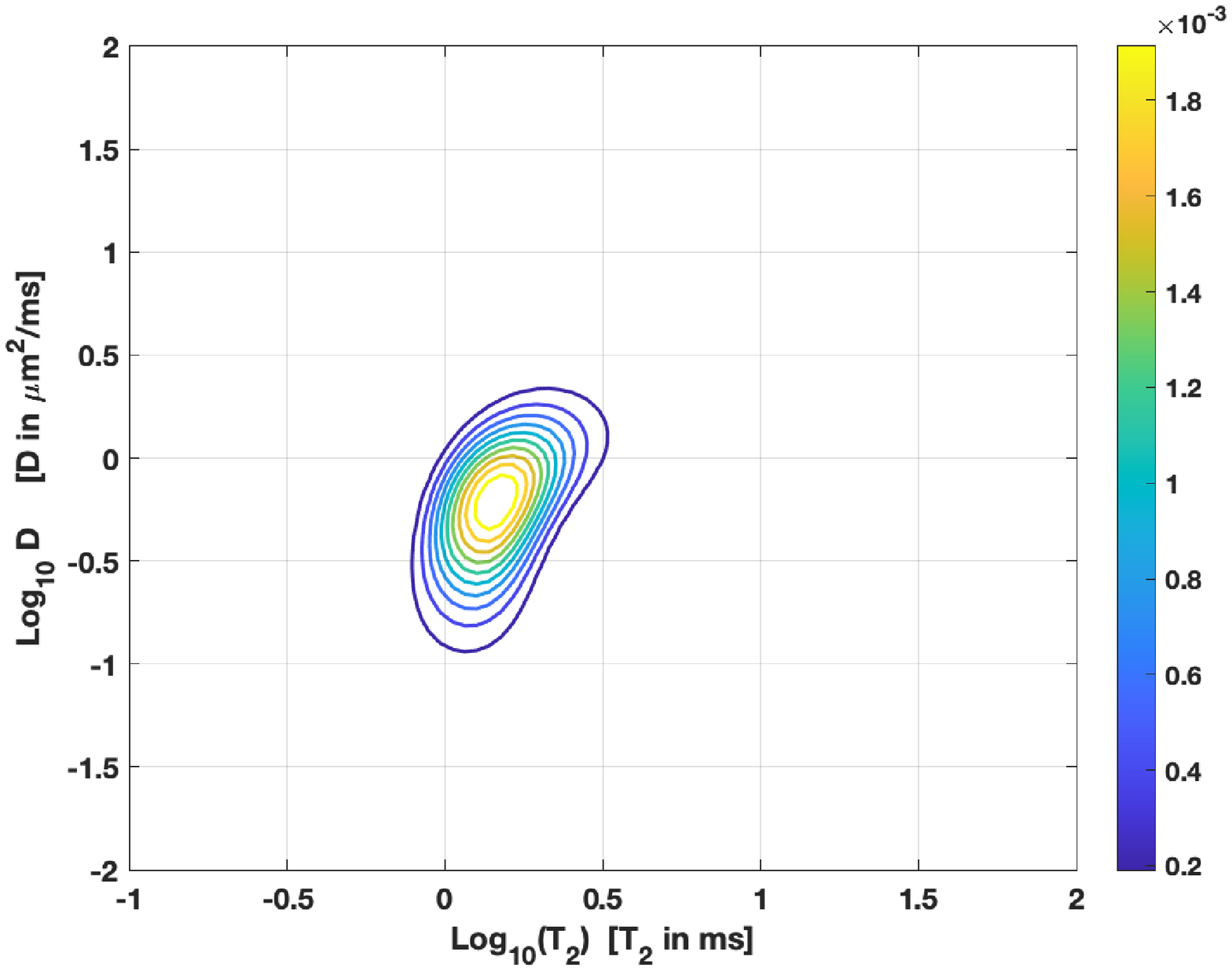}
    \includegraphics[scale=0.3]{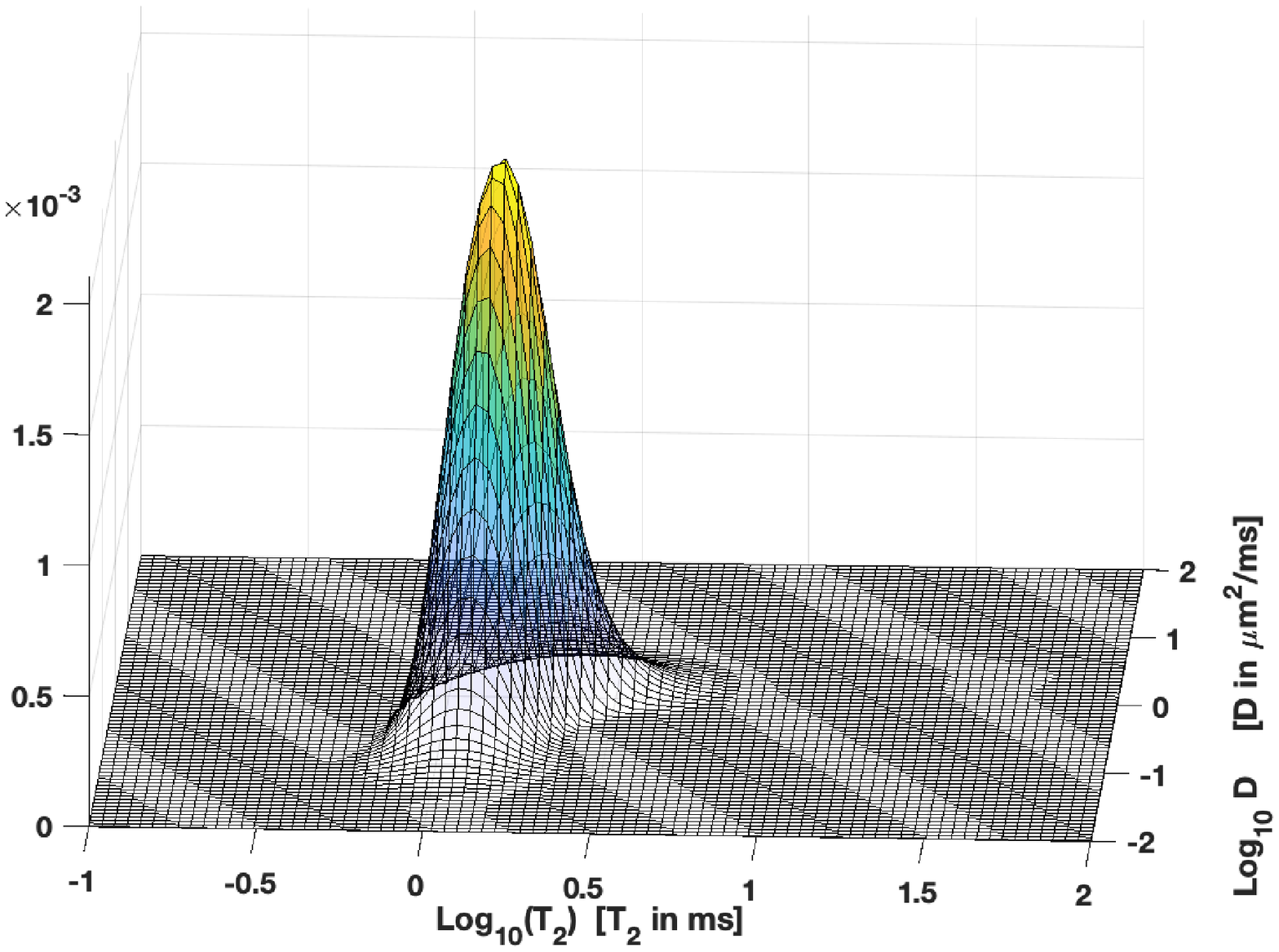} \\
    (a) \hspace{2cm} (b)
    \caption{(a) Contour map  and (b) relaxation map computed by \Name .}
    \label{fig:D_T2_1}
\end{figure}

\begin{figure}[h!]
    \centering
    \includegraphics[scale=0.3]{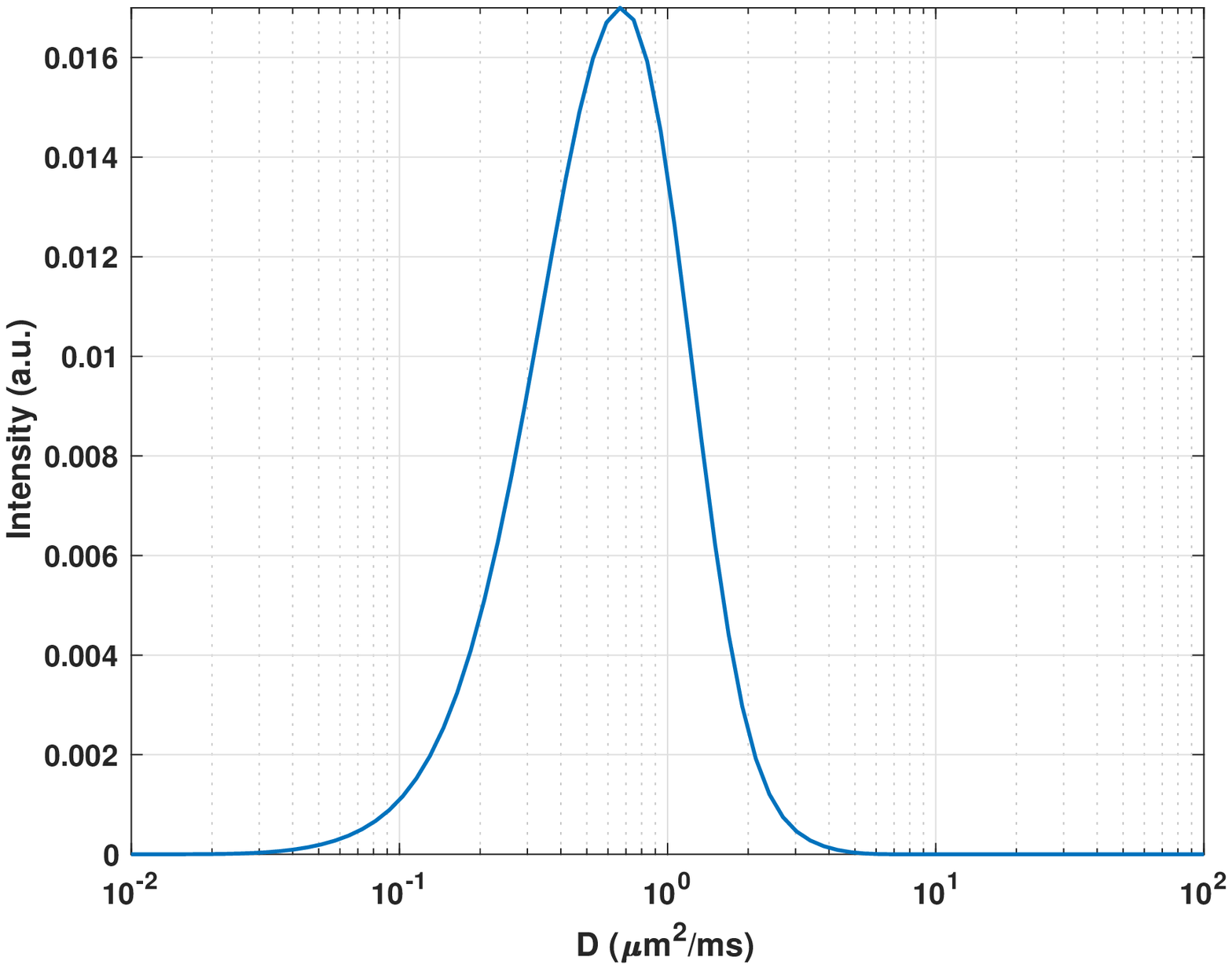}
    \includegraphics[scale=0.3]{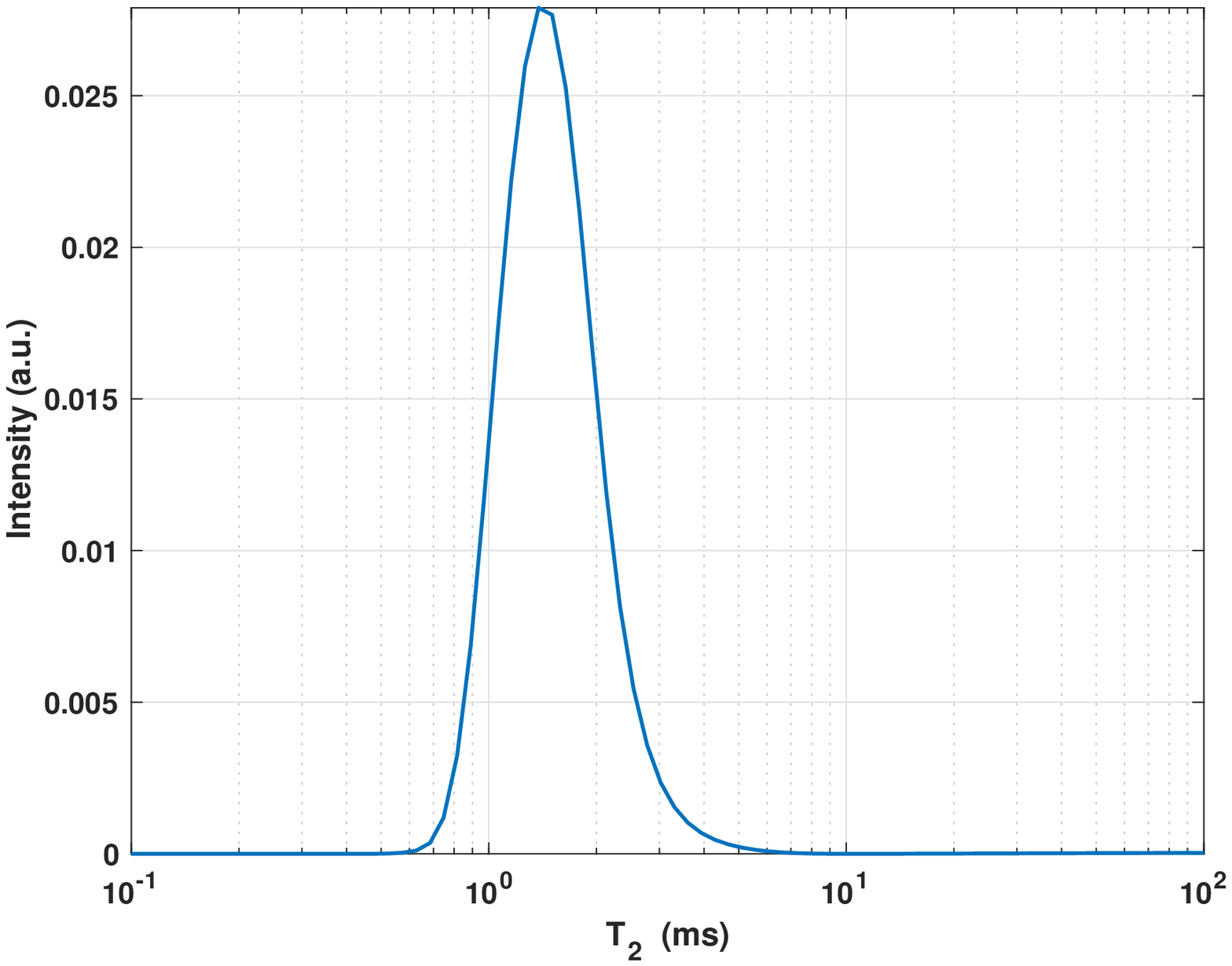} \\
    (a) \hspace{2cm} (b)
    \caption{$D$ one dimension projection (a) and $T_2$ one dimension projection (b).}
    \label{fig:D_T2_2}
\end{figure}

Figure \ref{fig:D_T2_1} shows the contour  and the relaxation map computed by \Name.
The projections on the diffusion $D$ and $T2$ directions are shown in figure \ref{fig:D_T2_2}, where is shown a single $T_2$ peak at about 10 ms, which probably corresponds to capillary water. With a TE of 400 $\mu$s shorter peaks (for example interlayer water) are not detected.

Finally, information about the reconstruction quality can be obtained by the post-processing step. 
\begin{figure}[h!]
    \centering
    \includegraphics[scale=0.3]{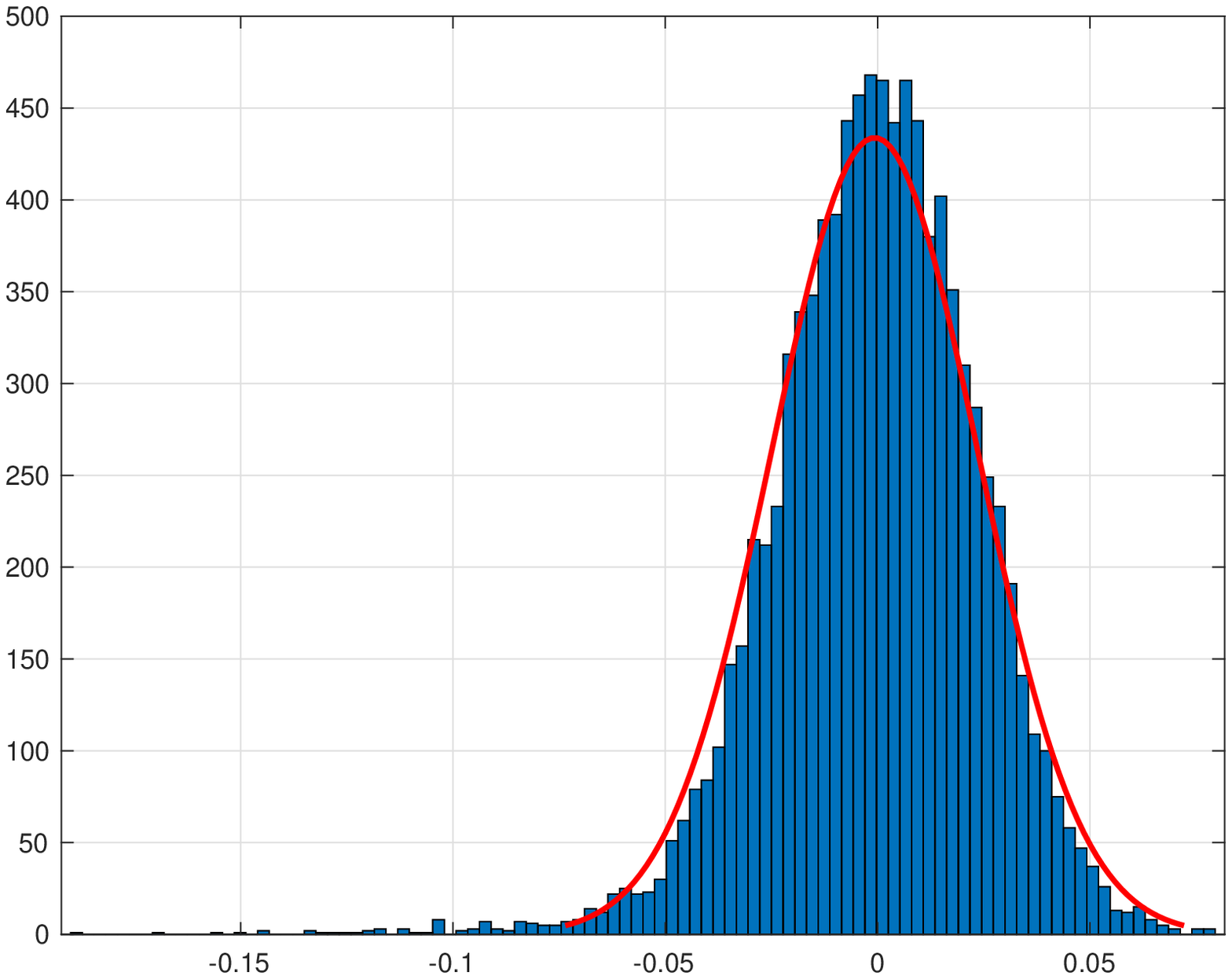}
    \includegraphics[scale=0.3]{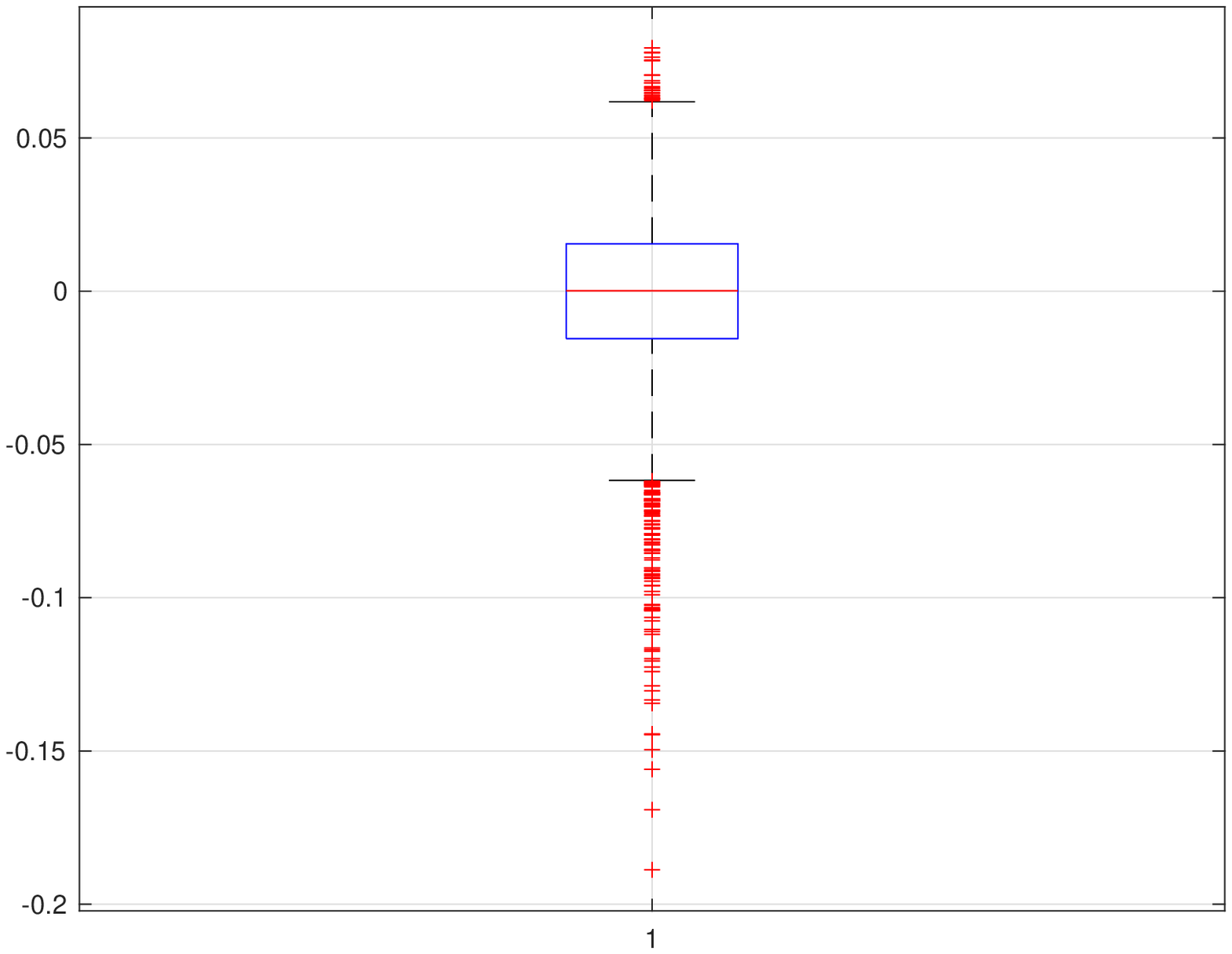} \\
    (a) \hspace{5cm} (b)
    \caption{Post-Processing. (a) Histogram of Residual values (blue bars) and Fitted Normal distribution  (red line) (b) Box Plot:
    red horizontal line represents the median value ($1.7224 \ 10^{-4})$;
    bottom and top edges of the box contain the $25$th ($ -1.5454 \ 10^{-2}$) and $75$th ($ 1.5496 \ 10^{-2}$) percentiles;
    red   '+' marker represents the outliers.}
    \label{fig:D_T2_3}
\end{figure}
The histogram of the residual distribution in figure \ref{fig:D_T2_3}(a) shows a good agreement with the normal distribution with variance $\sigma=0.024$ and mean $\mu=-6.9121 \ 10^{-4}$  represented in red line.
Information about the lack of symmetry (skewness) and weight of the distribution tails (kurtosis) is contained in the structure {\tt out\_data}, i.e.:
$$\hbox{{\tt out\_data.skewness}} = -0.6265, \ \ \ \hbox{{\tt out\_data.kurtosis}}= 5.4128.$$

Following \cite{hair2010},  data is considered to be normal since  skewness is between $-2$ and $+2$ and kurtosis is between $-7$ and $+7$.
The box-plot in figure \ref{fig:D_T2_3}(b) shows the median ($1.7224 \ 10^{-4})$ in the central mark, and 
the $25$th ($ -1.5454 \ 10^{-2}$) and $75$th ($ 1.5496 \ 10^{-2}$) percentiles at the bottom and top edges of the box containing $9400$ out of $9600$ points ($97.9 \%$). The  outliers,  amounting  to $150$ points, are plotted individually using red   '+' marker symbols. The remaining $50$ points lie  in the intervals $[-0.06347, -0.0155]$ and $[0.0155,0.061]$, represented by the lower and upper whiskers and the box lower and upper boundaries.
The file {\tt Parameters.txt} in the {\tt output} subfolder contains the following information about input tolerances, data size,  algorithm iterations and computation time.
\begin{verbatim}
----------------------------------------------------------
MUpen2D Input Parameters 
 MUpen2D tol = 1.000000E-04,
 Projected Gradient Tol = 1.000000E-03 
 SVD Threshold = 1E-16 
 Data size = 48 x 80  
----------------------------------------------------------
Number of Inversion channels:  horizontal 80, vertical  80 
Final Relative Residual Norm = 3.6393E-01 
Total MUpen2D Iterations = 28
Total FISTA Iterations = 50361 
Computation Time = 15.61428 s
----------------------------------------------------------
\end{verbatim}
\section*{Conclusion}
This algorithm paper describes the \Name\ open source software implementing the method proposed in \cite{ComputGeoSci2021} for the inversion of 2D NMR data.   Moreover   several representative examples are presented in detail  to help the interested users to include their own data. We believe that \Name\ can be  usefully applied in different applications whenever a phenomenon is modelled as an  exponentially decaying function.
\section*{Acknowledgments}
This work was partially supported by the Istituto Nazionale di Alta Matematica,
Gruppo Nazionale per il Calcolo Scientifico (INdAM-GNCS).

%
%
%

\end{document}